\documentclass[aip,jcp,reprint,floatfix,showkeys,a4paper]{revtex4-1}

\usepackage{hyperref}
\usepackage{inputenc}
\inputencoding{utf8}
\usepackage[OT1]{fontenc}
\usepackage{siunitx}
\usepackage{graphicx}
\usepackage{amssymb,amsmath}
\usepackage{commath}
\usepackage{amsmath}
\usepackage{amsthm}
\usepackage{amsfonts}
\usepackage{amssymb}
\usepackage[danish,english]{babel}
\usepackage{color}
\usepackage{tikz}
\usepackage[version=3]{mhchem}

\begin{document}
\selectlanguage{english}


\title{Isochronal superposition and density scaling of the $\alpha$-relaxation from pico- to millisecond}
\author{Henriette Wase Hansen}
\affiliation{Glass and Time, IMFUFA, Department of Science and Environment, Roskilde University, Postbox 260, DK-4000 Roskilde, Denmark}
\affiliation{Institut Laue-Langevin, 71 avenue des Martyrs,
CS 20156, 38042 Grenoble Cedex 9, France}
\author{Bernhard Frick}
\affiliation{Institut Laue-Langevin, 71 avenue des Martyrs,
CS 20156, 38042 Grenoble Cedex 9, France}
\author{Simone Capaccioli}
\affiliation{Dipartimento di Fisica, Universit{\`a} di Pisa, Largo B. Pontecorvo 3, I-56127 Pisa, Italy}
\author{Alejandro Sanz}
\author{Kristine Niss} \email{kniss@ruc.dk}
\affiliation{Glass and Time, IMFUFA, Department of Science and Environment, Roskilde University, Postbox 260, DK-4000 Roskilde, Denmark}

\date{\today}


\begin{abstract}

  The relaxation dynamics in two van der Waals bonded and one hydrogen-bonding
  molecular liquids is studied as a function of pressure and
  temperature by incoherent neutron scattering using simultaneous
  dielectric spectroscopy. The dynamics is studied in a range of alpha
  relaxation times from nano- to milliseconds, primarily in the
  equilibrium liquid state. In this range we find that isochronal
  superposition and density scaling work not only for the two van
  der Waals liquids, but also for the hydrogen-bonding liquid, though
  the density scaling exponent is much smaller for the latter. Density scaling
  and isochronal superposition are seen to break down for
  intra-molecular dynamics when it is separated in time from the alpha
  relaxation in close agreement with previous observations from
  molecular dynamics simulations.

\end{abstract}


\maketitle


\section{Introduction}
A full understanding of the dynamics in liquids is still a pending challenge in chemical physics
research\cite{Berthier11,Ediger12}. A better understanding of the dynamics has implications
not only for small organic molecular liquids as will be presented in
this paper, but also for other materials such as polymers, proteins
and metals\cite{Arbe12,Khodadadi15,Luo17}. The dynamics in liquids spans many orders of timescales from the equilibrium liquid above the melting temperature, where relaxation times are shorter than nanoseconds, to the viscous slowing down in the supercooled state as the glass transition is approached and the $\alpha$-relaxation reaches \SI{100}{\s}.

Slowing down of the liquid dynamics and the glass transition are
traditionally observed by cooling a liquid, but they are also seen upon
compression. New insight into liquid dynamics has been gained in the past decades by
introducing pressure as an additional variable to temperature \cite{Roland05}.

T{\"o}lle et al. \cite{Tolle01} observed
isochronal superposition of neutron spectra on
picosecond timescales. Isochronal superposition is the
observation of an invariance of the spectral shape and linewidth along lines in the
temperature-pressure phase diagram of constant
relaxation time\cite{Frick03_IS}. The observation of isochronal superposition implies that the spectral shape is determined by the relaxation time or that they are controlled by the same underlying mechanism\cite{Ngai05}.

Density scaling is another experimental observation from pressure
experiments, based on the ideas from the soft repulsive potential\cite{Tolle01,Dreyfus}.
It was shown for several systems that the
viscosity or the relaxation time as a function of temperature and
pressure could be brought to collapse when expressed in terms of the
parameter $\Gamma=\rho^\gamma/T$ (Refs.~\onlinecite{Casalini04_DS,Alba-Simionesco04,Roland05}), where $\gamma$
is a material specific constant. Density scaling has since been shown
to apply to numerous different systems and implies
that in these systems the relaxation time is governed by the parameter $\Gamma$.

A complete understanding of the dynamics in liquids must encompass
pressure behaviour, and therefore also the observations of density
scaling and isochronal superposition from pressure experiments. One
proposed approach is to assume the existence of isomorphs as suggested
by isomorph theory \cite{Gnan09,Dyre14}. The basic idea is that two thermodynamic
state points are isomorphic if they have the same potential energy
landscape upon scaling. This gives rise to curves in the phase diagram
along which dynamics on all timescales presented in reduced dimensionless units are invariant. The dynamics in this view is on all
timescales determined by the same control parameter, $\Gamma$, which
in the simplest case is given by $\Gamma=\rho^\gamma/T$. Density
scaling and isochronal superposition are therefore natural
consequences of isomorphs. A conjecture of isomorph theory is that it
only works for $R$-simple systems, i.e. systems that in molecular
dynamics simulations show a strong correlation between the potential
energy and the virial
\cite{Bailey08I,Pedersen10,Ingebrigtsen12a,Dyre06}. From an
experimental point of view, the conjecture is that the idea of
$R$-simple systems is manifested in systems without directional
bonding or competing interactions. Moreover, isomorph theory only
describes inter-molecular interactions and to be certain to
exclude intra-molecular modes we consider experimental candidates for
simple liquids to be those with no large secondary relaxations.

From dielectric spectroscopy, it has previously been shown that
isochronal superposition works better for van der Waals liquids than
for hydrogen-bonding liquids \cite{Roed13}, although several other
studies also with dielectrics\cite{Hensel04,Grzybowska06,Adrjanowicz16,Puosi16} and molecular dynamics simulations \cite{Romanini17} have shown that density scaling and isochronal superposition
apply to hydrogen-bonding systems as well.

In a recent paper, we showed that the picosecond dynamics along the glass transition as a function of temperature and pressure, i.e. fast relaxations and vibrations that are completely separated from the
$\alpha$-relaxation, measured with neutron spectroscopy were invariant
for two simple van der Waals liquids\cite{hwh18}.
For these simple liquids, we observed isochronal superposition of the dynamics
separated by 14 orders of magnitude, which was interpreted as a genuine signature of these liquids having isomorphs. The implication of this finding is that for simple
liquids with isomorphs, the phase diagram can be reduced to
just one dimension with the control parameter $\Gamma$, in contrast to
an experimentalist's normal temperature-pressure phase diagram.
A hydrogen-bonding liquid with directional bonding was chosen to test a more complex system in contrast to the simple van der Waals liquids. We showed how the relation between picosecond dynamics and the glass transition broke down for the hydrogen-bonding complex liquid, in agreement with the predictions from isomorph theory.

Most previous experimental studies of glass-forming liquids, also the ones at short
timescales as probed for example by neutron spectroscopy, take place in
the viscous or supercooled liquid, thus, in the proximity of the glass
transition. In this paper, we focus on the dynamics in the liquid
state with fast $\alpha$-relaxation times. We expand the test of
density scaling and isochronal superposition to experimental timescales in the equilibrium liquid when the $\alpha$-relaxation is on the same timescale as neutron spectroscopy. Timescales that also coincides with those accessible with computer simulations.

We apply for the first time systematic inelastic fixed window
pressure studies from the backscattering instrument IN16B at the
Institut Laue-Langevin and use those to test density scaling.
Most of the data presented in this paper have been measured in a
newly developed pressure cell for performing simultaneous dielectric and
neutron spectroscopy as a function of temperature and pressure \cite{Sanz18}, thereby
expanding the accessible neutron scattering to include longer timescales, while assuring that the data are collected under exactly the same experimental conditions.

In this paper, we present scaling of the $\alpha$-relaxation on pico-
to millisecond timescales for three different systems, two simple van
der Waals liquids and a more complex hydrogen-bonding sample, studied with neutron and dielectric spectroscopy. We begin by introducing experimental details in the next section. Data are
presented in the Sec.~III containing three subsections, one for
each sample, followed by the discussion in Sec.~IV.


\section{Experimental details}

\begin{figure}[htpb!]
\centering
\includegraphics[width=0.25\columnwidth]{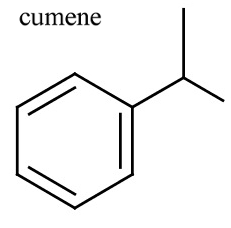} \hspace{1 em}
\includegraphics[width=0.6\columnwidth]{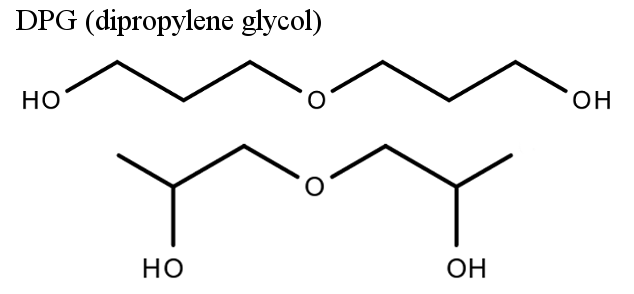} \\
\includegraphics[width=0.9\columnwidth]{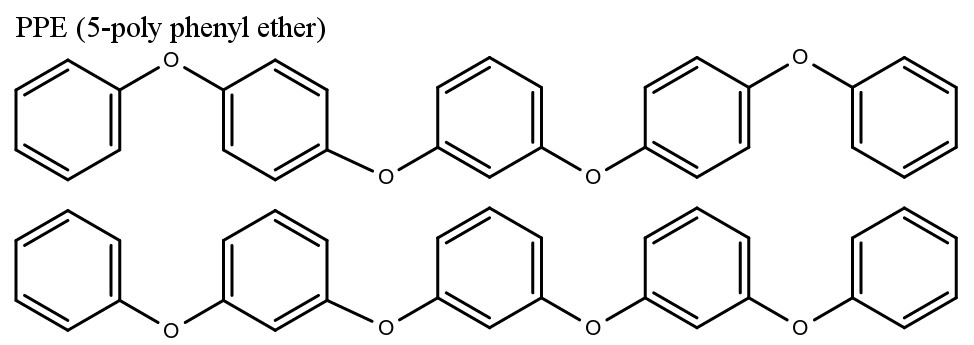}
\caption{Sketches of the molecular structure of the three samples studied in this chapter. DPG and PPE are mixtures of isomers.}\label{fig:exp:molecules}
\end{figure}

The three samples presented in this paper are two van der Waals
liquids, isopropyl benzene (cumene) and 5-polyphenyl ether (PPE), and
the hydrogen-bonding liquid dipropylene glycol (DPG)
(Fig.~\ref{fig:exp:molecules}). Cumene and DPG were purchased from
Sigma Aldrich and PPE from Santolubes, and all three samples were used as
acquired. The two van der Waals liquids are considered simple in the
sense that they do not possess pronounced secondary relaxations or
intra-molecular motion visible on either subnanosecond timescales in the neutron signal or in
the dielectric signal \cite{hwh17}, and have previously been shown to have behaviour in agreement with predictions from isomorph theory \cite{hwh18,Xiao15,Roed13}. The hydrogen-bonding liquid DPG represents a
complex system with directional bonding, a clear excess wing visible in the dielectric
signal \cite{Casalini04} and methyl-group rotation observable on nanosecond timescale
with neutron spectroscopy, and has a high pressure response for a hydrogen-bonding liquid. The molecules are sketched in
Fig.~\ref{fig:exp:molecules} and details about the samples can be
found in Table~\ref{table:exp:details}.

\begin{table}
\centering
\begin{tabular}{l l l l l l}
 & $T_g(P_\mathrm{amb})$ \hspace{2 em}  & $m$ \hspace{2 em} & $\gamma$ \hspace{2 em} & d$T_g$/d$P$ \hspace{1 em} & Refs.\\
 & \hspace{1 em} [K] & & & [K/MPa] & \\
 \hline
cumene 	& \hspace{1 em} 126 & 70 & 4.8 & 0.085 & \cite{Ransom17,hwh18}\\
PPE 	& \hspace{1 em} 245 & 80 & 5.5 & 0.18\phantom{0} & \cite{Hecksher13,Gundermann13_thesis,hwh18} \\
DPG 	& \hspace{1 em} 195 & 65 & 1.5 & 0.075 & \cite{Grzybowska06,Grzybowski06_gamma,hwh18} \\
\end{tabular}
\caption{Details on the three samples: cumene (isopropyl benzene), PPE (5-polyphenyl ether) and DPG (dipropylene glycol). The fragility, $m$, is a measure of the super-Arrhenius behaviour introduced by Angell \cite{Angell91}: $m=\frac{\mathrm{d} \log_{10}\tau(T)}{\mathrm{d}(T_g/T)}\biggr\rvert_{T_g}$.}\label{table:exp:details}
\end{table}

The data presented in this paper have been measured at the Institut
Laue-Langevin (ILL) at the backscattering instrument IN16B  and IN13 and the
time-of-flight instrument IN5 \cite{Exp-6-05-961,Exp-LTP-6-7}. The timescale of the neutron
experiments is determined by the energy resolution and the energy transfer of the neutron
instruments; the coarser the energy resolution, the faster the timescale. We measured cumene and DPG on IN16B, accessing nanosecond dynamics with the standard Si(111) setting with
$\lambda=\SI{6.271}{\angstrom}$ in the $Q$-range from 0.4 to
\SI{1.8}{\per\angstrom} with an energy resolution of $\sim\SI{0.75}{\micro\eV}$ corresponding to nanosecond timescales in the energy range $\pm\SI{30}{\micro\eV}$. The elastic intensity of cumene was also measured at the backscattering instrument IN13 with an energy resolution of \SI{8}{\micro\eV} in the $Q$-range $0.2-\SI{4.9}{\per\angstrom}$, i.e. an order of magnitude faster than IN16B.
Cumene and PPE were both measured on IN5 with a wavelength of \SI{5}{\angstrom}
with an energy resolution of \SI{0.1}{\milli\eV} corresponding to a timescale of
$\sim\SI{10}{\pico\s}$, i.e. two orders of magnitude faster than IN16B, in the $Q$-range from $1.2-\SI{1.9}{\per\angstrom}$ binned in steps of $\SI{0.1}{\per\angstrom}$ in the energy range $\pm\SI{2}{\milli\eV}$ . Full spectra of cumene were also measured at IN5 with wavelength \SI{8}{\angstrom} with energy resolution \SI{0.015}{\milli\eV} accessing a timescale of roughly \SI{0.1}{\nano\s} in the $Q$-range from
$0.5-\SI{1.2}{\per\angstrom}$ in energy range $\pm\SI{1}{\milli\eV}$.
All data reduction, subtraction of background, normalisation to monitor and vanadium, corrections for
cell geometry and instrument efficiency, have been done in
LAMP\cite{LAMP}. Except for the IN16B data on cumene, all data were
measured in a newly developed pressure cell for doing simultaneous
dielectric and neutron spectroscopy under high pressure
\cite{Sanz18}. The temperature range for the high-pressure cell is $2-\SI{320}{\K}$ and is controlled by the standard orange cryostat available at the ILL. The pressure range is $0.1-\SI{500}{\mega\pascal}$. Both the temperature and pressure range are determined by the aluminium alloy of the high-pressure cell, which is used because of its low background in the neutron signal. This new sample cell is a unique tool for combining techniques, thereby accessing dynamics on different timescales with high precision, making sure that data are collected under exactly the same sample conditions.

We utilise fast switching between the elastic and inelastic fixed window
scans unique for the backscattering instrument IN16B (Ref.~\onlinecite{Frick12}). These
scans offer a fast overview of how a
parameter such as temperature or pressure in the full $Q$-range
affects the dynamics. The IFWS data presented in this paper were
measured with an energy offset at \SI{2}{\micro\eV} and a maximum is
observed when the spectral weight at the offset energy is highest
as a relaxational process passes through the instrument
window, for example caused by an increase of the quasielastic broadening. The nature of the dynamics, whether it is a local process or of translational character, can be determined from the $Q$-dependence. That is, if a relaxation shows change in the $Q$-dependence as a function of temperature or pressure it is a delocalised translational motion, whereas a local process will be independent of $Q$ as a function of pressure or temperature \cite{Bee88}. All spectra are shown summed over $Q$ for better statistics since we generally observe the same trend for all values of $Q$, unless otherwise stated. More data and details on samples and data are available in Ref.~\onlinecite{hwh_thesis}.

Equations of state to calculate the density at various state points
for cumene is from Ref.~\onlinecite{Ransom17}, for PPE from
Ref.~\onlinecite{Gundermann13_thesis}, and for DPG from
Ref.~\onlinecite{Grzybowski11}. The scaling exponents used for testing
density scaling for the three samples were found in previous
publications: Ransom et al. \cite{Ransom17} measured the glass
transition temperature of cumene in a large pressure range, up to
$\sim\SI{4}{\giga\pascal}$, and found $\gamma=4.8$, and used this to
density scale light scattering data\cite{Li95} and viscosity
data\cite{Barlow66,Ling68}. The scaling exponent for PPE was found
from dielectric data to be $\gamma=5.5$ (Ref.~\onlinecite{Gundermann13_thesis}) and used
in scaling of dielectric spectra\cite{Xiao15}. The two van der Waals
liquids complement each other in terms of experimental accessibility:
cumene has a low glass transition temperature which means that the
$\alpha$-relaxation reaches subnanoseconds, probed by our scattering experiments,
within the $(T,P)$-range accessible by our sample cell, and furthermore, cumene crystallizes easily in
the range between the melting point and the glass transition.
PPE, on the other hand, does not crystallize and we can therefore easily
study dynamics in the viscous dynamic range,
whereas the $\alpha$-relaxation is too slow within the $(T,P)$-limits of our equipment compared
to the neutron timescales. The hydrogen-bonding liquid DPG was found to have a scaling
exponent of $\gamma=1.5$ (Ref.~\onlinecite{Grzybowski06_gamma}), a quite low value of $\gamma$
compared to the two van der Waals liquids, but typical for
hydrogen-bonding liquids \cite{Casalini04_DS,Roland05,Niss06}.
This scaling exponent allows for scaling of the $\alpha$-relaxation in the range from microseconds to milliseconds \cite{Grzybowska06}, while it does not provide a satisfactory collapse for dynamics close to the glass transition.

As mentioned earlier, according to isomorph theory, the relevant quantities to investigate are in reduced units, which are per definition dimensionless (presented with a tilde). Examples of length and time units, $l_0$ and $t_0$, respectively, used in this paper are:
\begin{equation}
\begin{aligned}
l_0 & = \rho^{-1/3} \\
t_0 & = \rho^{-1/3}(m_p/k_\mathrm{B}T)^{1/2},
\end{aligned}
\end{equation}
where $m_p$ is the average particle mass and $\rho$ is the number density. For length, wave vector and frequency, respectively, the reduced dimensionless units denoted with a tilde are then given by,
\begin{equation}\label{eq:exp:reducedunits}
\begin{aligned}
\tilde{\mathbf{r}} & = \mathbf{r}/l_0 = \mathbf{r} \rho^{1/3} \\
\tilde{{Q}} & = {Q}~l_0 = Q\rho^{-1/3}  \\
\tilde{\omega} & = \omega~t_0 = \omega~\rho^{-1/3}(m_p/k_\mathrm{B}T)^{1/2}.
\end{aligned}
\end{equation}
The frequency unit $\omega$ multiplied by $\hbar$ corresponds to the energy transfer. We assume the average particle mass is constant and set $m_p=k_\mathrm{B}=1$. Effectively, the reduced unit for energy transfer thus becomes
\begin{equation}\label{eq:exp:energy_red}
\tilde{\omega} = \omega\rho^{-1/3}T^{-1/2},
\end{equation}
where $\rho$ is now the volumetric mass density. In the dynamic range investigated in this paper, the change in density is in the percent range, thus, the inverse of the cubic root of the density will only have a small effect on the scaling of the energy transfer. Plotting data in reduced energy units will therefore mainly be affected by temperature changes. The scaling has a visible effect only for higher energy transfer, and we will therefore ignore the reduced units for the low frequency IFWS in the next section. Expressing $Q$ in reduced units will result in changes of around 1\%, which will be within the uncertainty of the data and is therefore also neglected.


\section{Results}

In this section, we present high-pressure neutron scattering data from the two van der Waals liquids, cumene and PPE, and the hydrogen-bonding liquid, DPG, presented in three subsections according to sample.

\subsection{van der Waals liquid -- cumene}

\begin{figure}[htpb!]
\centering
\includegraphics[width=0.49\columnwidth]{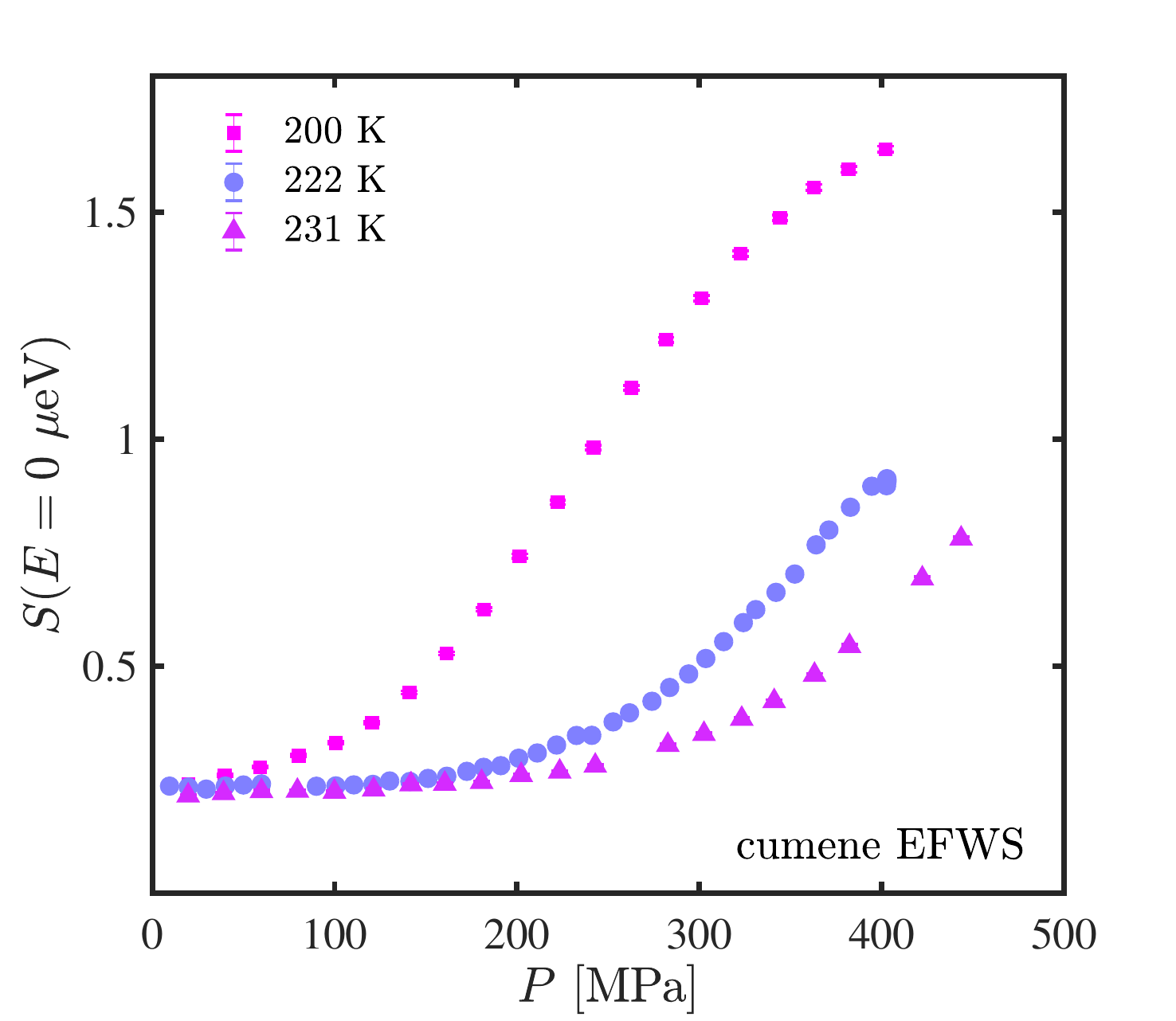}
\includegraphics[width=0.49\columnwidth]{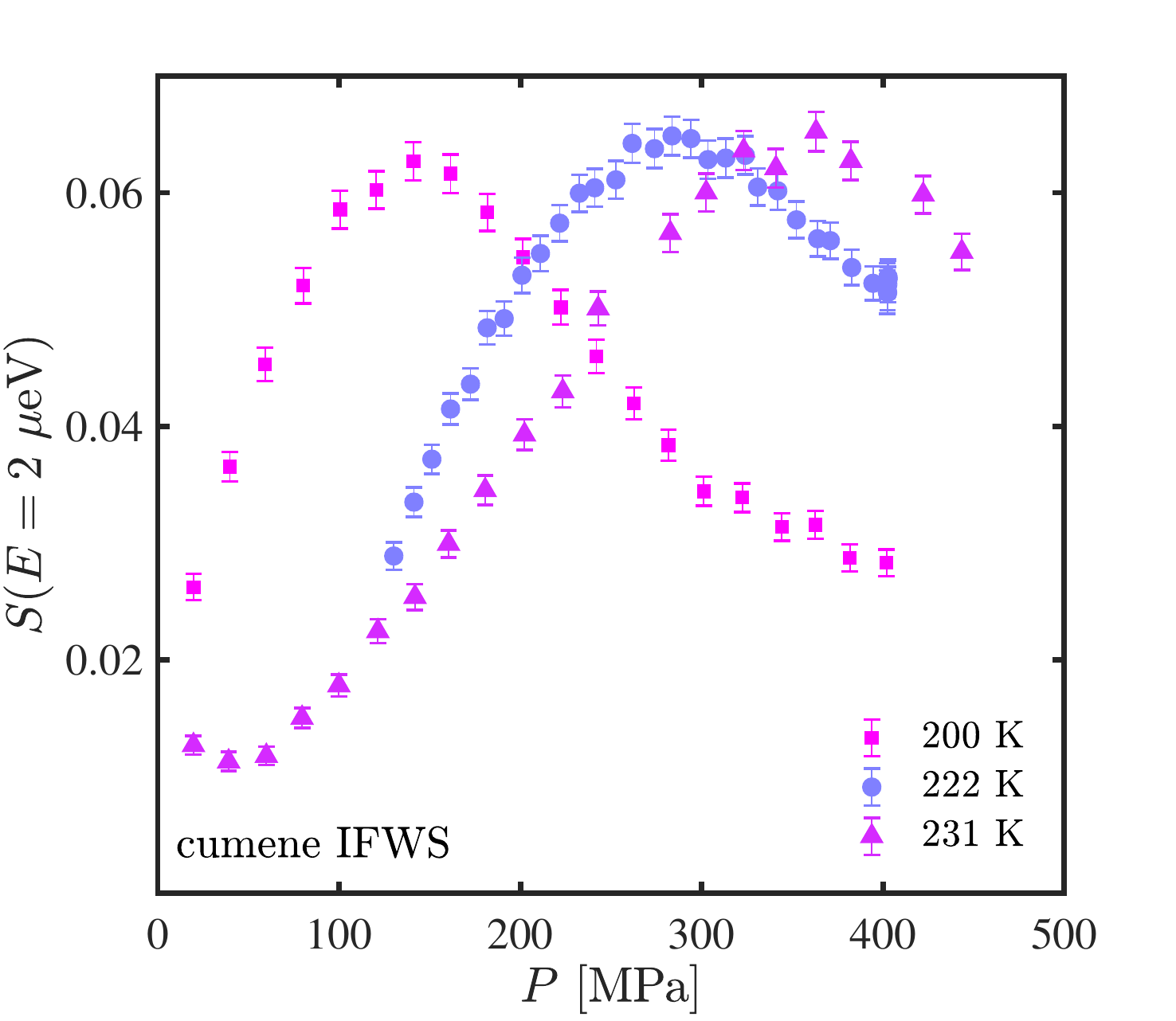} \\
\includegraphics[width=0.49\columnwidth]{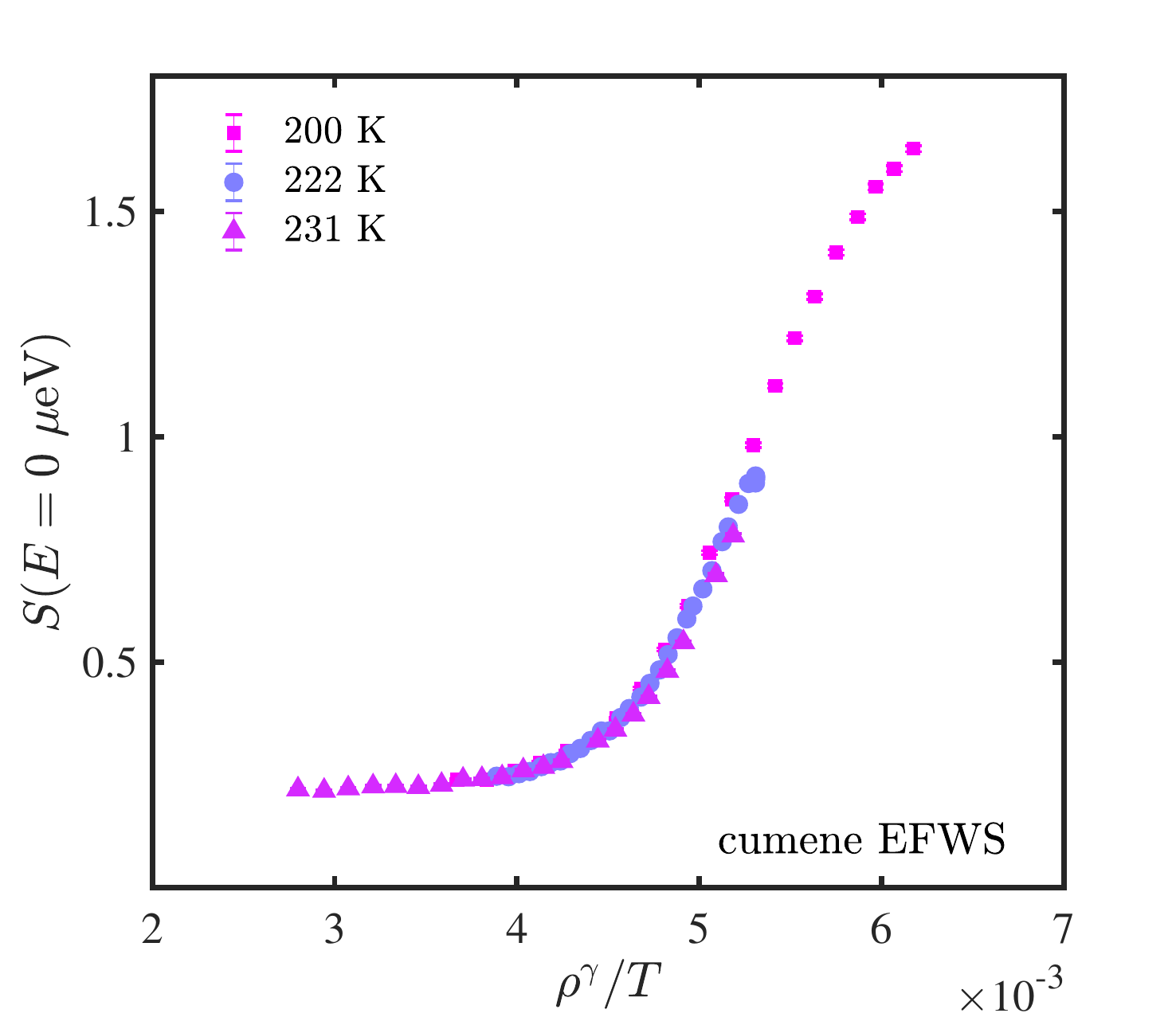}
\includegraphics[width=0.49\columnwidth]{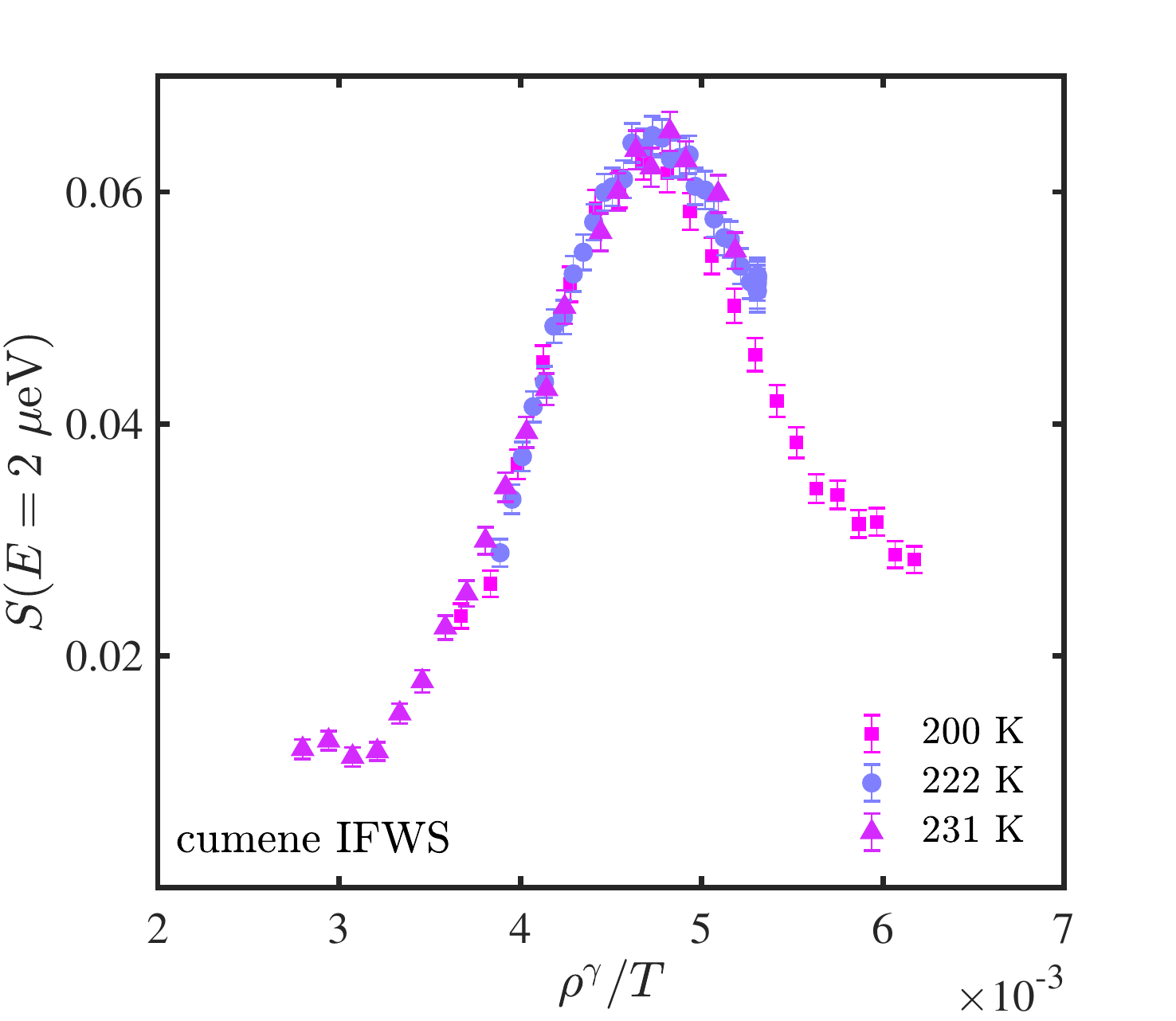}\\
\caption{EFWS (left) and IFWS (right) from IN16B on cumene summed over $Q$. Intensity of EFWS and IFWS plotted for three isotherms (\SI{200}{\K}, \SI{222}{\K}, \SI{231}{\K}) as a function of pressure (top) and as a function of $\Gamma=\rho^\gamma/T$ with $\gamma=4.8$ (bottom).}\label{fig:res:cumene_fws}
\end{figure}

First, we present the fixed window scans from IN16B on cumene in Fig.~\ref{fig:res:cumene_fws}. In the top panel, the elastic and inelastic fixed energy window scan (EFWS and IFWS) intensity are shown as a function of pressure for three isotherms. From the EFWS ($\Delta E=\SI{0}{\micro\eV}$), the intensity is observed to increase on compression as the mobility decreases moving towards the glass transition. In the IFWS ($\Delta E=\SI{2}{\micro\eV}$), a maximum is observed in the intensity as a function of pressure as the $\alpha$-relaxation moves through the instrument window upon compression.

Density scaling predicts that the control parameter of the dynamics is $\Gamma=\rho^\gamma/T$ and we therefore plot the three isotherms from the FWS on IN16B as a function of $\rho^\gamma/T$ in the bottom panel of Fig.~\ref{fig:res:cumene_fws}, using the scaling exponent $\gamma=4.8$ determined by Ransom et al. \cite{Ransom17}. We observe how the data collapse to just one curve for both the EFWS and the IFWS. The same is observed for the elastic intensity from IN13, also a backscattering instrument, with an energy resolution that corresponds to a timescale one order of magnitude faster than IN16B. We are able to express the nanosecond dynamics along three isotherms in the equilibrium liquid with the $\alpha$-relaxation on nanosecond timescale as a function of the control parameter $\Gamma=\rho^\gamma/T$ using a $\gamma$ value found in the supercooled liquid at the glass transition. The scaling exponent, $\gamma$, is according to isomorph theory allowed to change with temperature and density and to depend on the dynamic range, and it has, of course, itself an uncertainty. With this type of experiment and uncertainty, we are not able to determine whether $\gamma$ is state-point dependent. In fact, the data can be made to collapse using values of $\gamma$ in the range from $4.2-4.8$. This type of analysis does not offer a precise way of determining $\gamma$.
However, within the experimental uncertainty that we have for this kind of experiment, e.g. knowing the exact temperature, we note how well the value found by Ransom et al. \cite{Ransom17} in the highly viscous liquid using completely different experimental techniques works here in the equilibrium liquid and therefore use $\gamma=4.8$ throughout this paper for cumene.

\begin{figure}[htpb!]
\centering
\includegraphics[width=0.49\columnwidth]{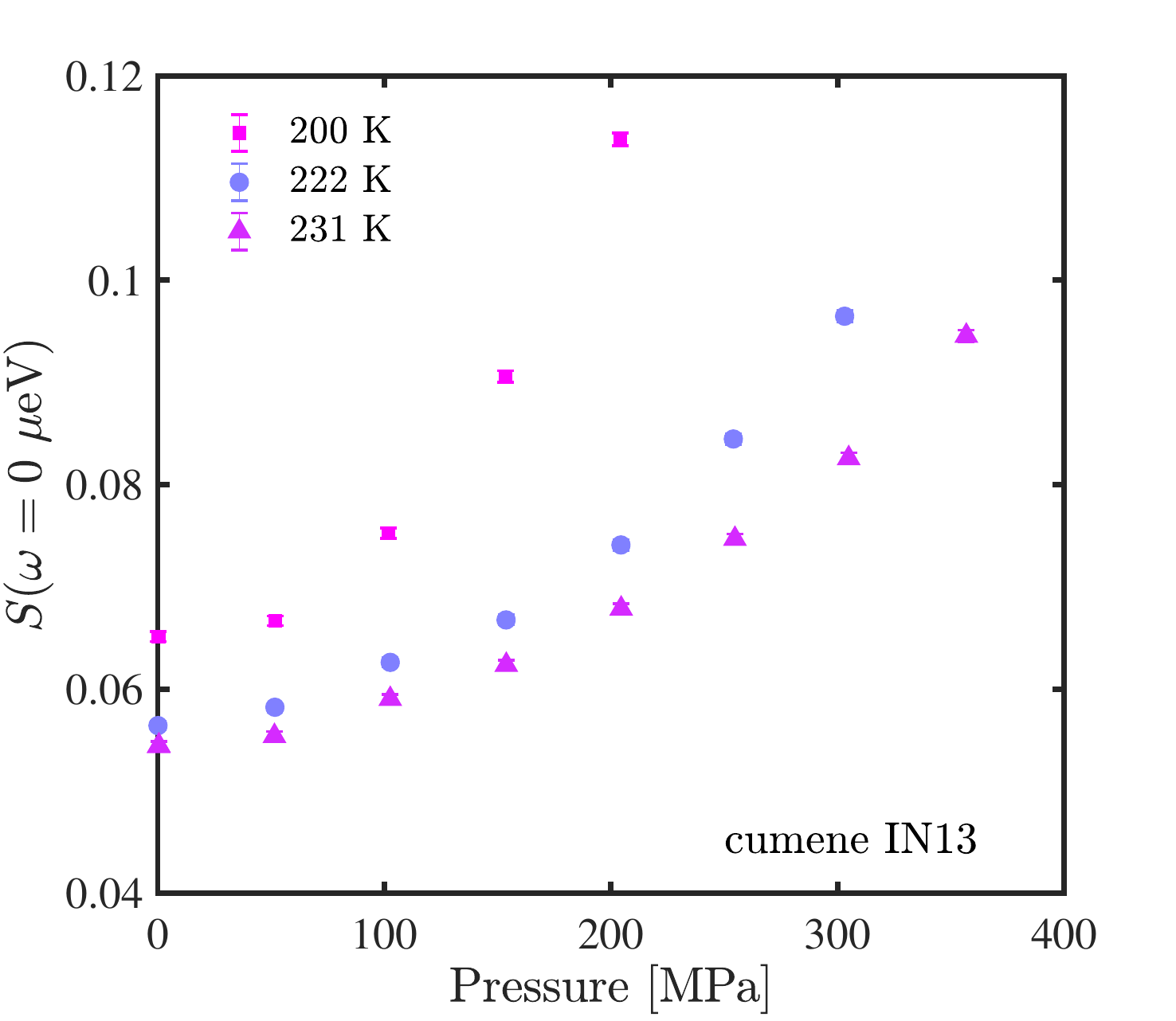}
\includegraphics[width=0.49\columnwidth]{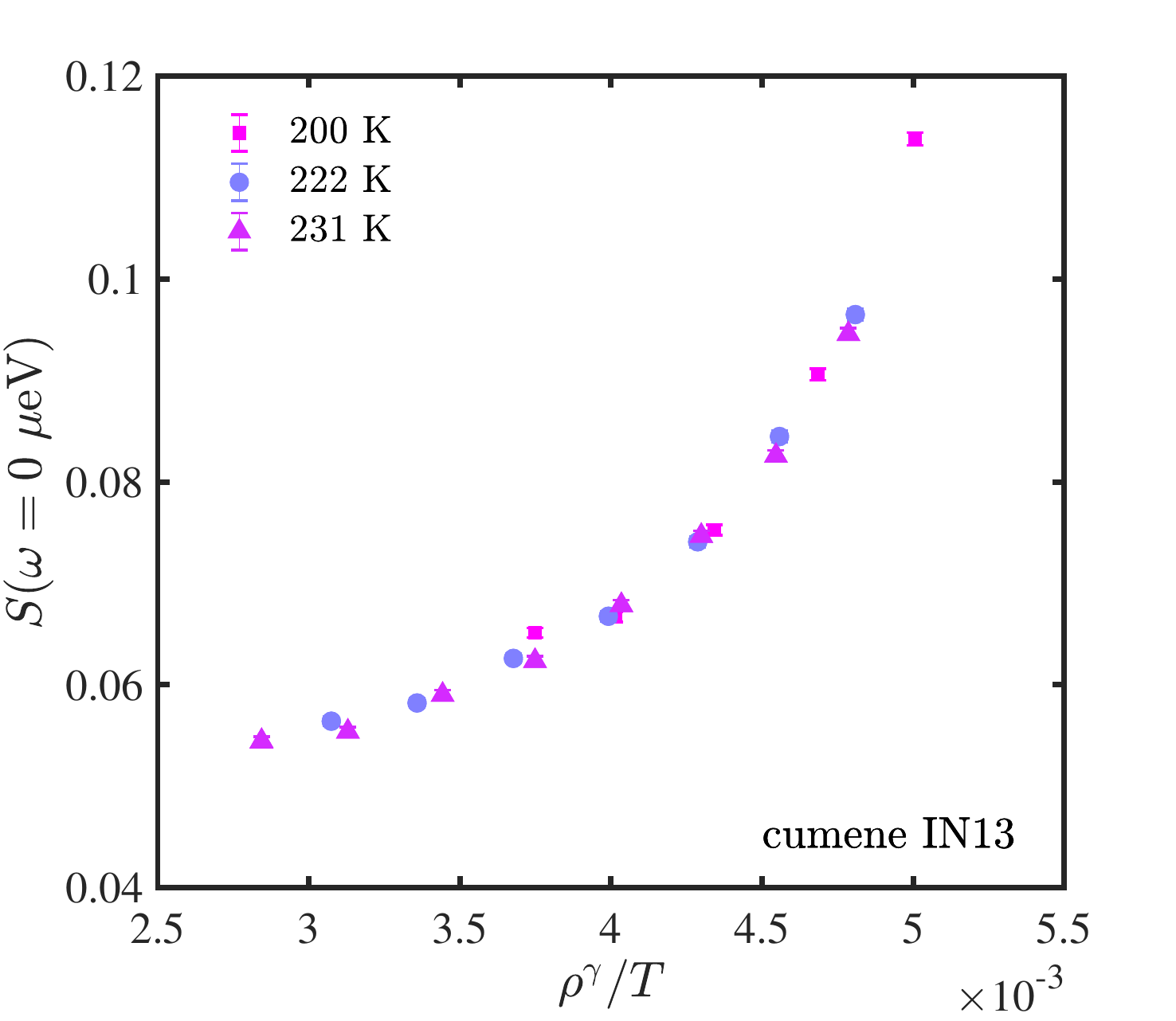}
\caption{Elastic intensity of cumene from IN13 summed over $Q$. Left: three isotherms (\SI{200}{\K}, \SI{222}{\K}, \SI{231}{\K}) as a function of pressure. Right: Density scaling of left plotted as a function of $\Gamma=\rho^\gamma/T$, $\gamma=4.8$.}\label{fig:res:cumene_in13}
\end{figure}

\begin{figure}[htpb!]
\centering
\begin{minipage}{0.3\columnwidth}
\vspace{-7 em}
{\noindent cumene \\
\vspace{0.5em} IN16B \\
\vspace{0.5em} $\sim\SI{e-9}{\s}$}
\end{minipage}
\includegraphics[width=0.32\columnwidth]{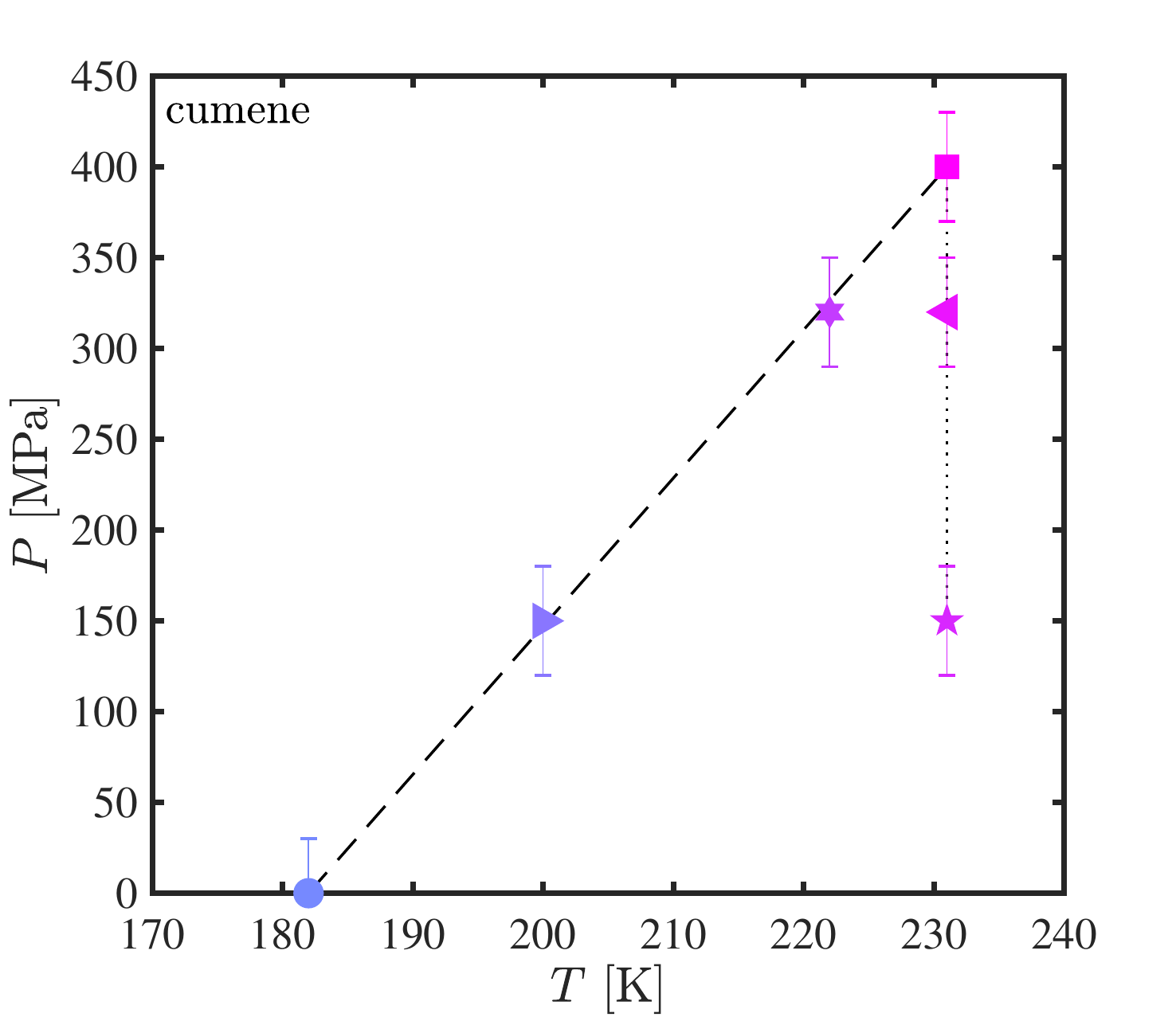} \\
\includegraphics[width=0.49\columnwidth]{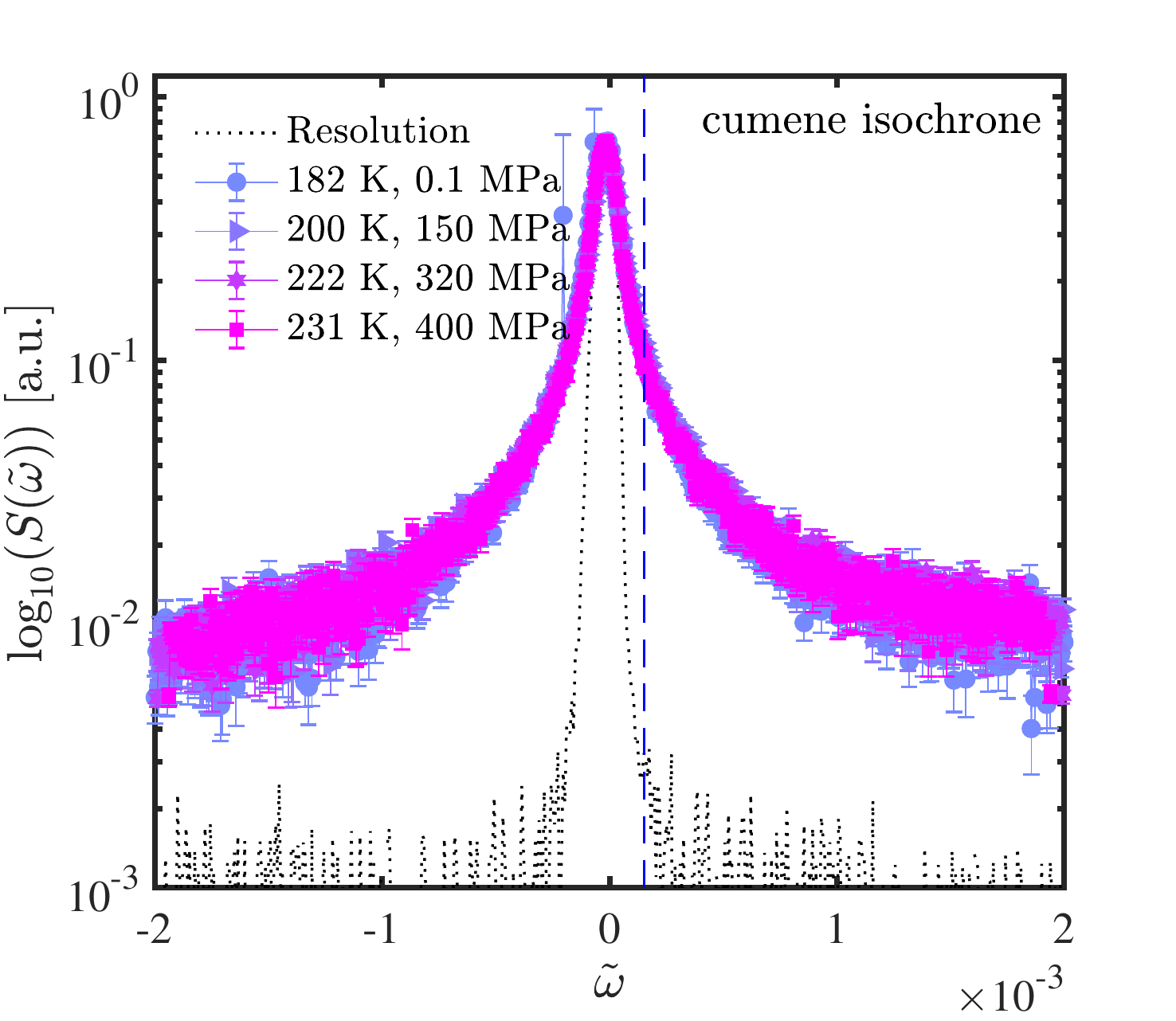} \includegraphics[width=0.49\columnwidth]{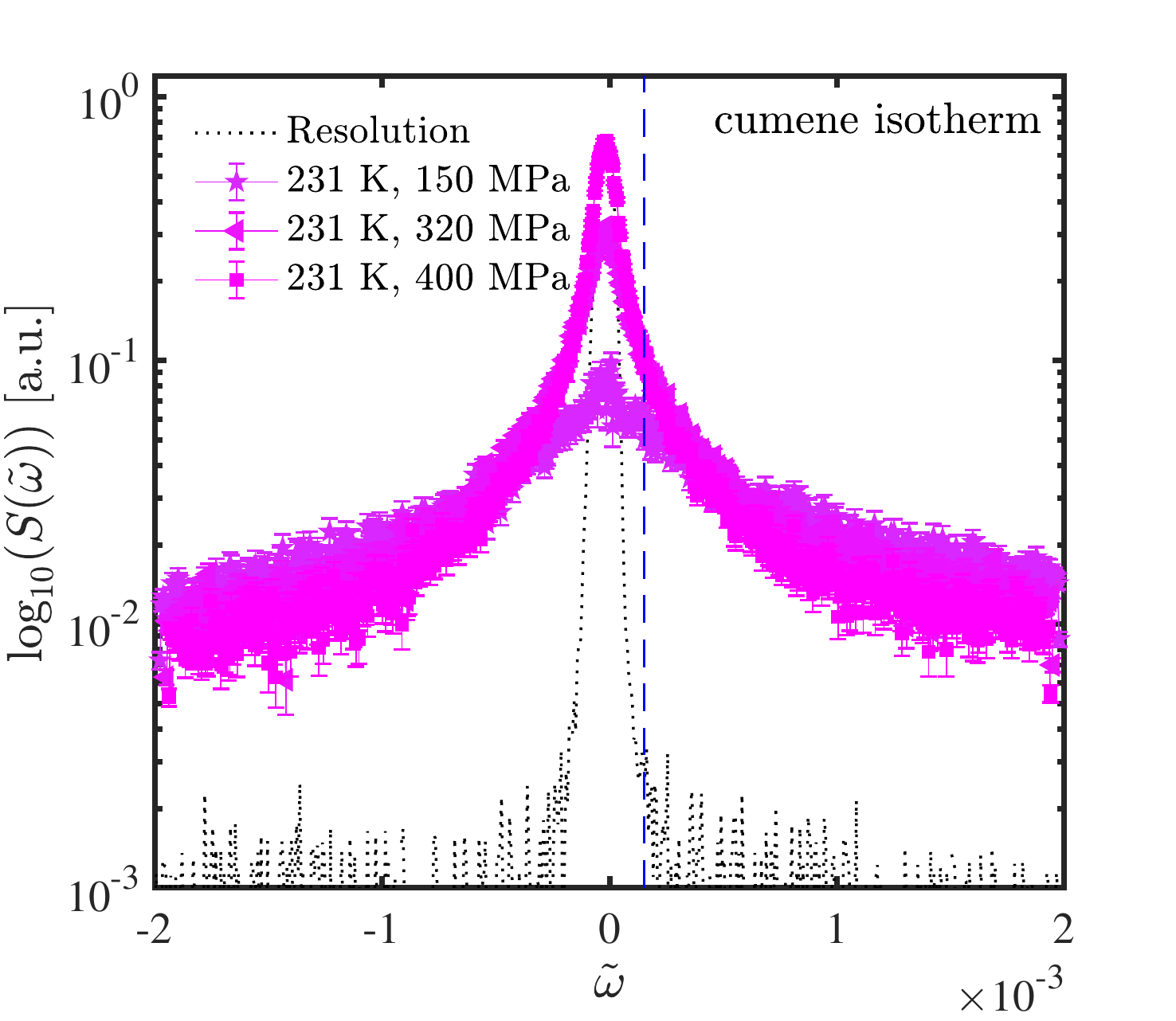}
\caption{Spectra of cumene from IN16B summed over $Q$. The $(T,P)$-phase diagram shows the studied state points for the isochrone as determined from fixed window scan on IN16B and an isotherm at \SI{231}{\K}. The blue dashed line in the spectra corresponds to an energy offset of \SI{2}{\micro\eV} used for the inelastic fixed window scans.}\label{fig:res:cumene_spectra}
\end{figure}

For the interpretation of the fixed window scans, we assume that by
knowing the intensity of the elastic peak ($\Delta
E=\SI{0}{\micro\eV}$) and the intensity at some offset energy, in this
case, $\Delta E=\SI{2}{\micro\eV}$, we have enough information to
interpret how the overall spectral shape and linewidth or quasielastic broadening changes with some variable,
in this case $T$ and $P$. Therefore, an indirect assumption of isochronal
superposition goes into this interpretation, i.e. we assume that
the linewidth of the spectral shape changes in a similar way for temperature and
pressure. This assumption can be tested by determining an isochrone
from the EFWS and IFWS of the three isotherms from IN16B by
identifying the same place in intensity on the curves for the three
isotherms and investigating the full spectra. In
Fig.~\ref{fig:res:cumene_spectra}, we present full spectra from IN16B
measured along the isochrone determined from the FWS and an
isotherm for comparison. The studied state points are shown in the $(T,P)$-phase
diagram. We observe an invariance of the spectral shape and quasielastic broadening along the
isochrone. For the isotherm, we observe less elastic intensity and more relaxation in the instrument window on decreasing pressure as the $\alpha$-relaxation time becomes faster on decreased pressure. This means that just one spectrum at a given $(T,P)$ gives the spectral shape and the FWS, which give the temperature and pressure dependence of the width, we are able to predict the spectral shape and linewidth along all state points that are isochronal to that spectrum.

\begin{figure}[htpb!]
\centering
\includegraphics[width=0.49\columnwidth]{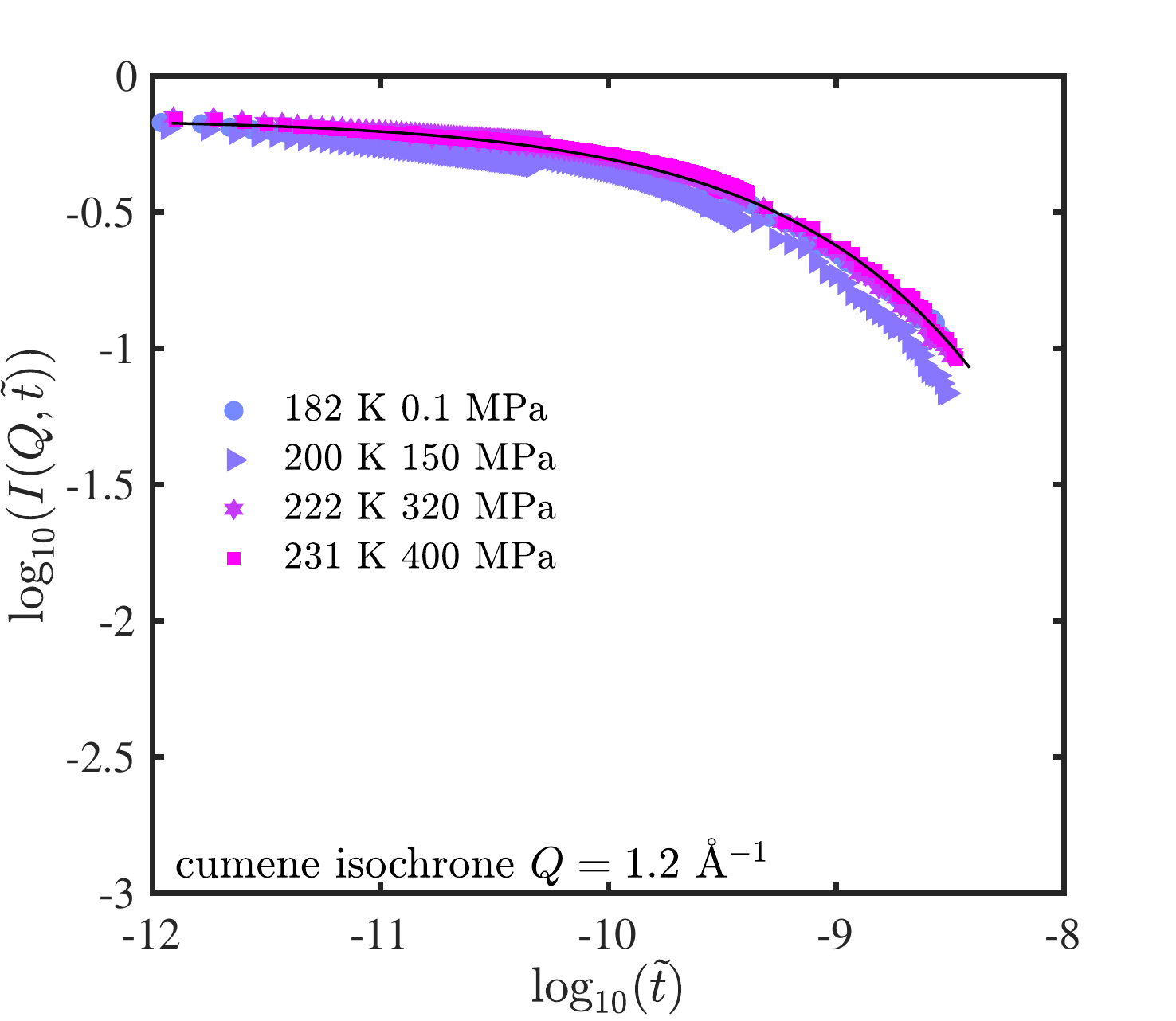}
\includegraphics[width=0.49\columnwidth]{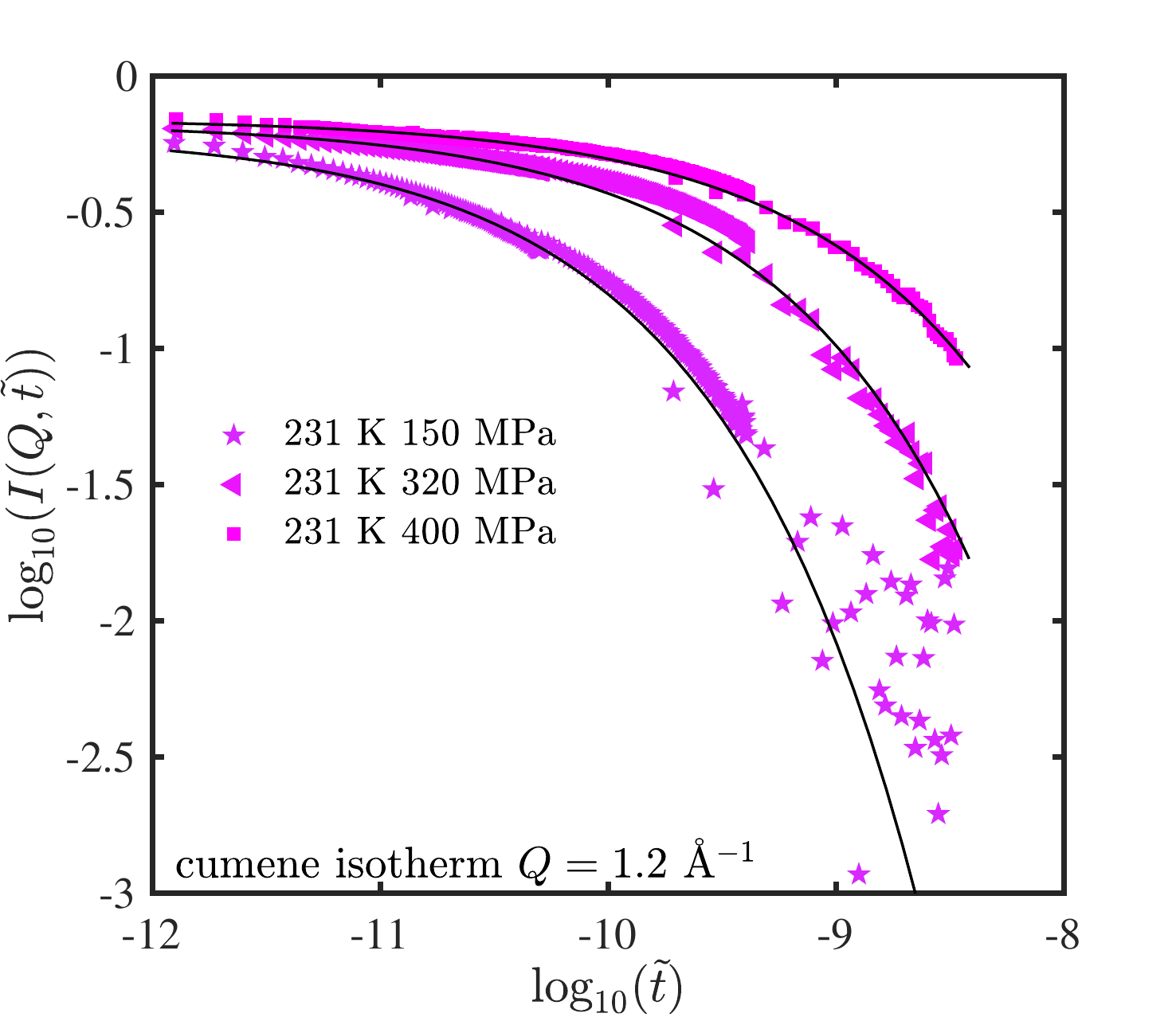}
\caption{Fourier transform of combined cumene spectra from IN16B and IN5 with $\lambda=\SI{5}{\angstrom}$ and \SI{8}{\angstrom} for $Q=\SI{1.2}{\per\angstrom}$. Full line is a stretched exponential fit to data from IN16B and IN5 with $\lambda=\SI{5}{\angstrom}$ and \SI{8}{\angstrom}.}\label{fig:res:cumene_fft}
\end{figure}

At the same state points as presented in Fig.~\ref{fig:res:cumene_spectra} for IN16B, spectra were measured on IN5 at two different resolutions, where we observed the same trend. In
Fig.~\ref{fig:res:cumene_fft}, we show the Fourier transformed combined spectra
of cumene from IN16B and IN5 with $\lambda=5$ and \SI{8}{\angstrom} at
$Q=\SI{1.2}{\per\angstrom}$. The Fourier transformed data have been corrected for resolution by merely dividing by the low-temperature spectrum of cumene for each spectrometer and wavelength. The total dynamic range spans close to four decades, and we observe a good degree of invariance for the
intermediate scattering funtion, $I(Q,t)$, along the isochrone determined from the
FWS on IN16B that are shown on the same scale as the isotherm for
comparison. For the isotherm, we observe as expected the relaxation time decreasing as pressure is released.

\begin{figure}[htpb!]
\centering
\includegraphics[width=0.49\columnwidth]{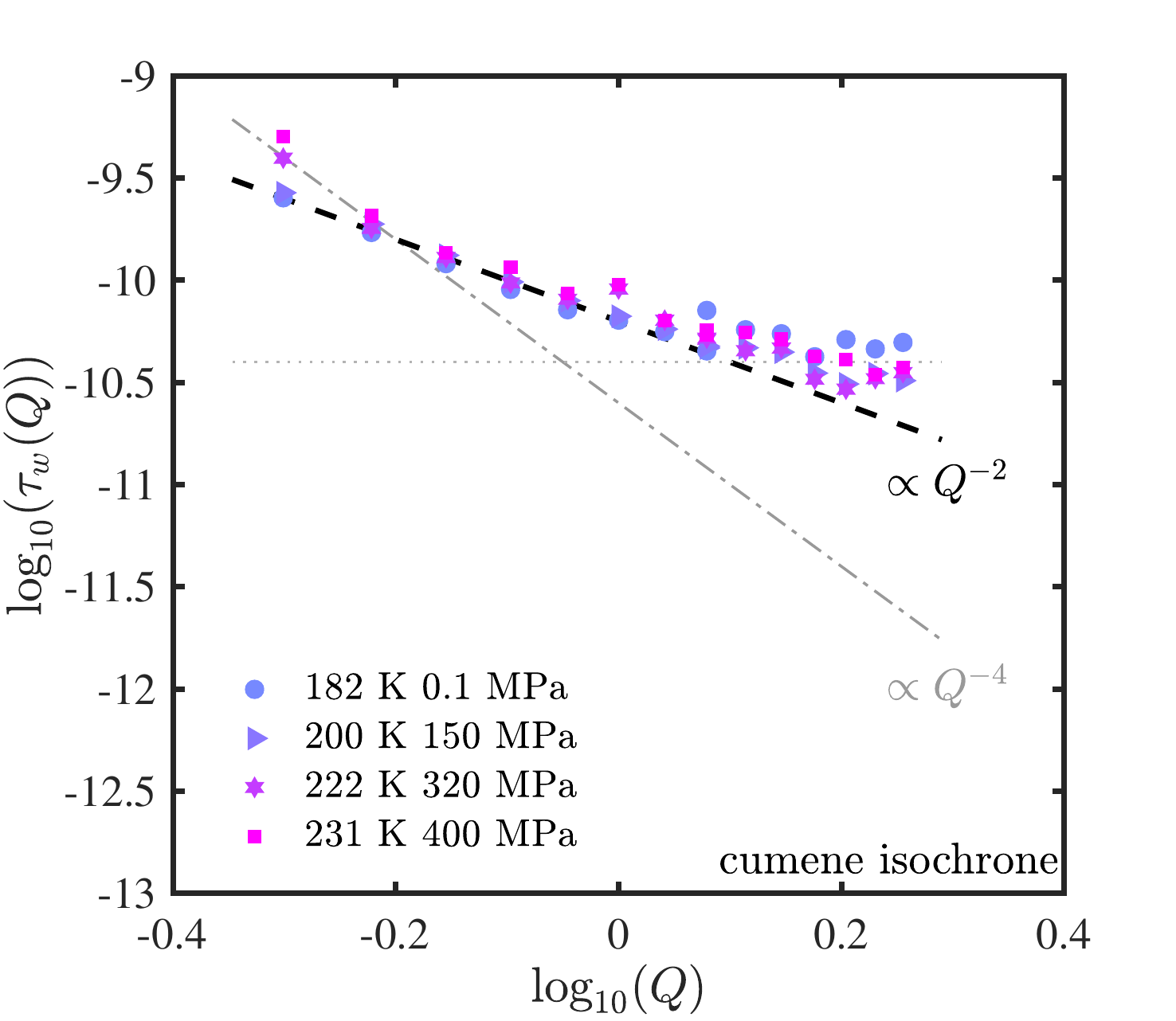}
\includegraphics[width=0.49\columnwidth]{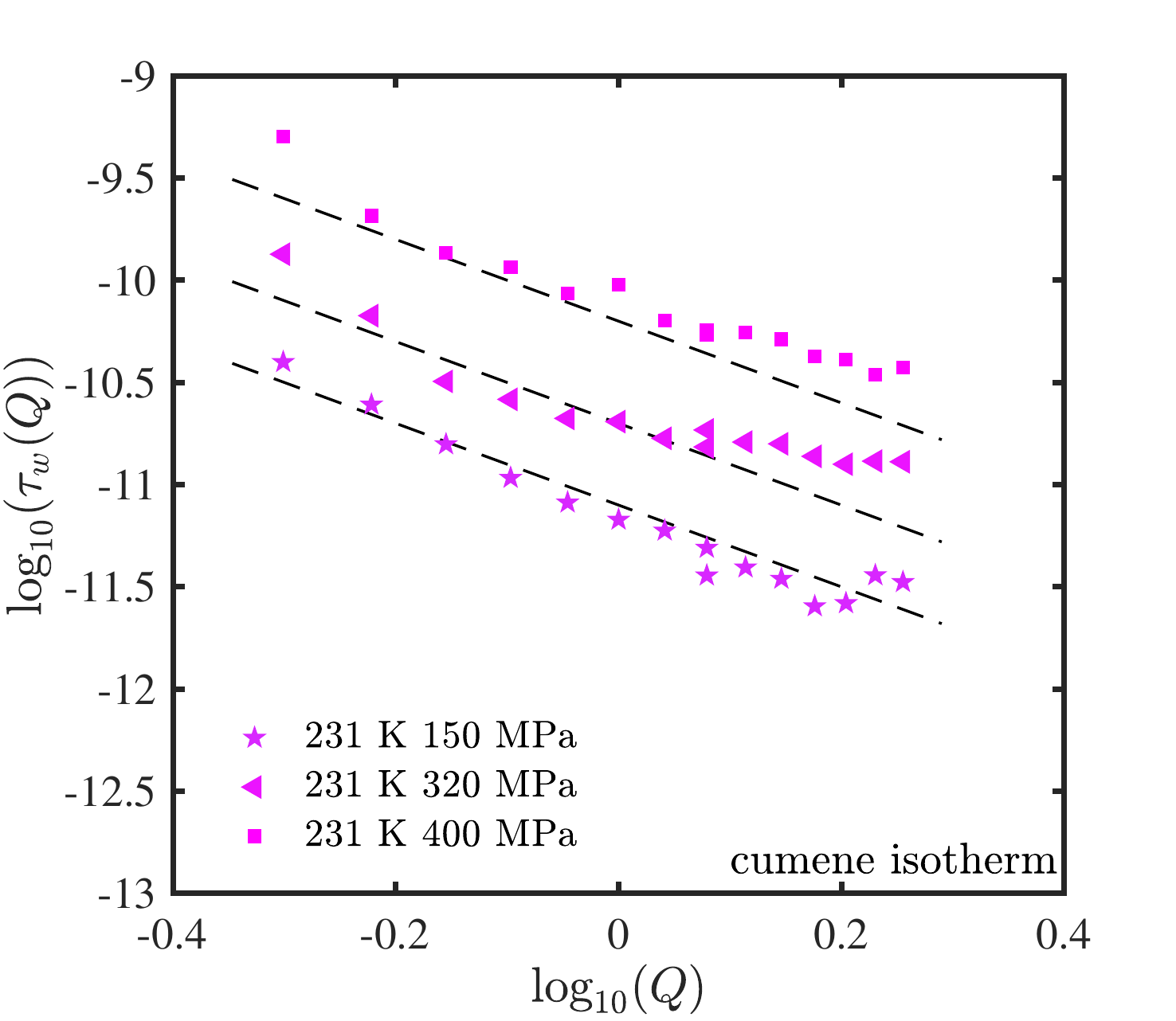}
\caption{The relaxation time $\tau_w$ as a function of $Q$ for cumene obtained from a stretched exponential fit to the Fourier transformed combined spectra from IN5 and IN16B along an isochrone and an isotherm. An example of the fit is shown as black lines in Fig.~\ref{fig:res:cumene_fft}. The dashed lines correspond to a $Q^{-2}$ dependence of the relaxation time.}\label{fig:res:cumene_fft_q}
\end{figure}

Relaxation processes in disordered systems are typically fitted with a stretched exponential as a phenomenological fitting function. We fitted the intermediate scattering functions with a stretched exponential as a function of $Q$:
\begin{equation}
I(Q,t)=Ae^{-\left(\frac{t}{\tau_w(Q)}\right)^\beta},
\end{equation}
where $A$ is the amplitude, $\tau_w$ is the $Q$-dependent relaxation time and $\beta$ is the stretching parameter. The stretching parameter was fixed at $\beta=0.5$ for all spectra as this gave the best fits, which indicates that the shape of all the spectra, independent of temperature and pressure, is the same, which means that isochronal superposition of the $\alpha$-relaxation becomes an evident observation.

The $Q$-dependence of the relaxation time is shown in Fig.~\ref{fig:res:cumene_fft_q} for the isochrone determined from the FWS on IN16B and for comparison, the $Q$-dependence of an isotherm. The relaxation time is observed to be fairly invariant along the isochrone, while the relaxation time is observed to slow down with increased pressure along the isotherm. For both the isochrone and the isotherm, the relaxation time is observed to become faster with increasing $Q$, showing a $Q^{-2}$-dependence in a large part of the accessible $Q$-range. This behaviour suggests a Gaussian behaviour which is observed to deviate for high $Q$, i.e. short distances, where the relaxation time is observed to flatten out away from the Gaussian behaviour. At low $Q$, there might be a slight tendency for the relaxation time to bend off into a $Q^{-2/\beta}$ behaviour as suggested in Ref.~\onlinecite{Colmenero93}, although this is not clear from these data sets. The $Q$-dependence observed for cumene suggests translational motion typical for structural relaxation.

\begin{figure}[htpb!]
\centering
\includegraphics[width=0.49\columnwidth]{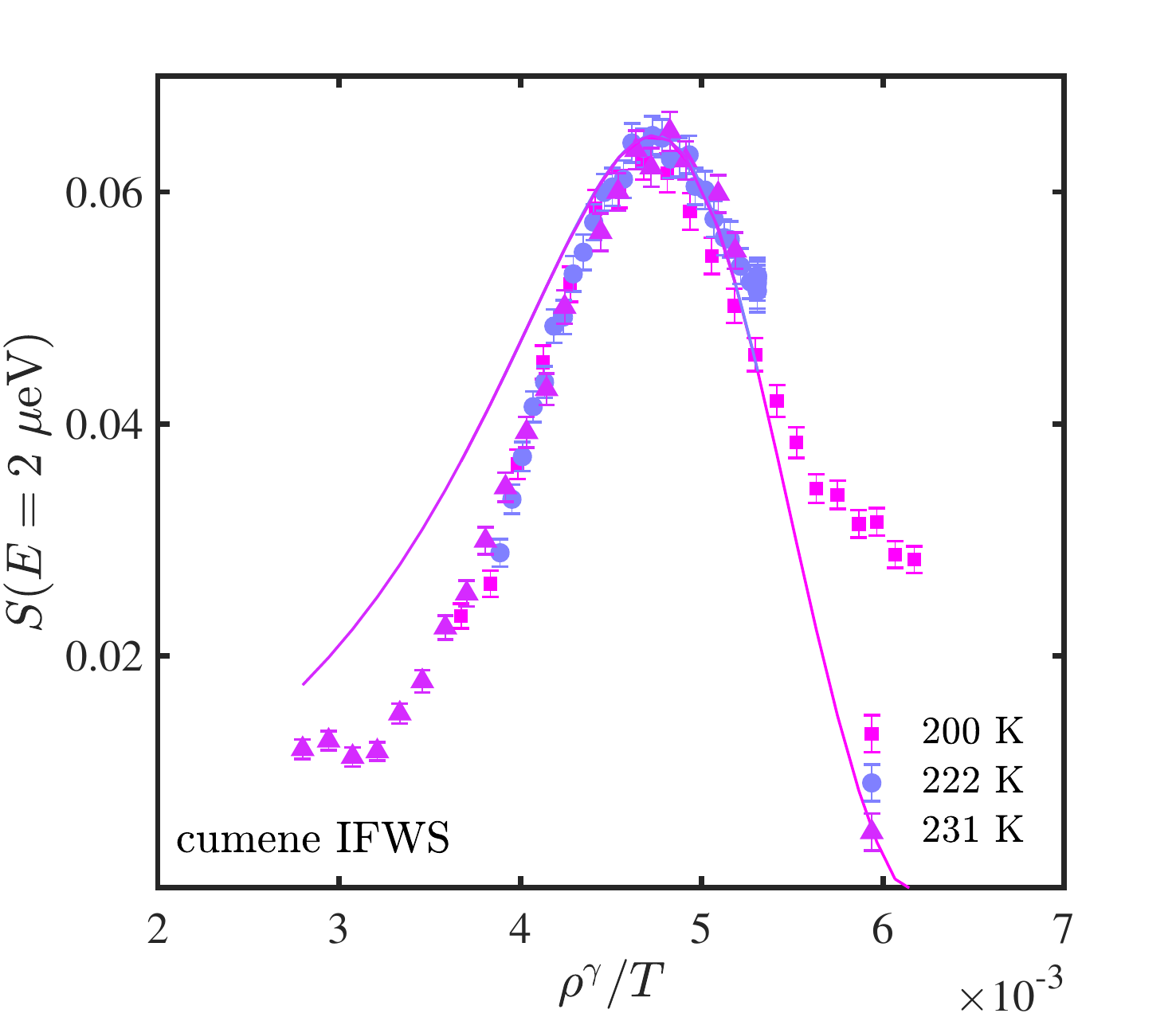}
\includegraphics[width=0.49\columnwidth]{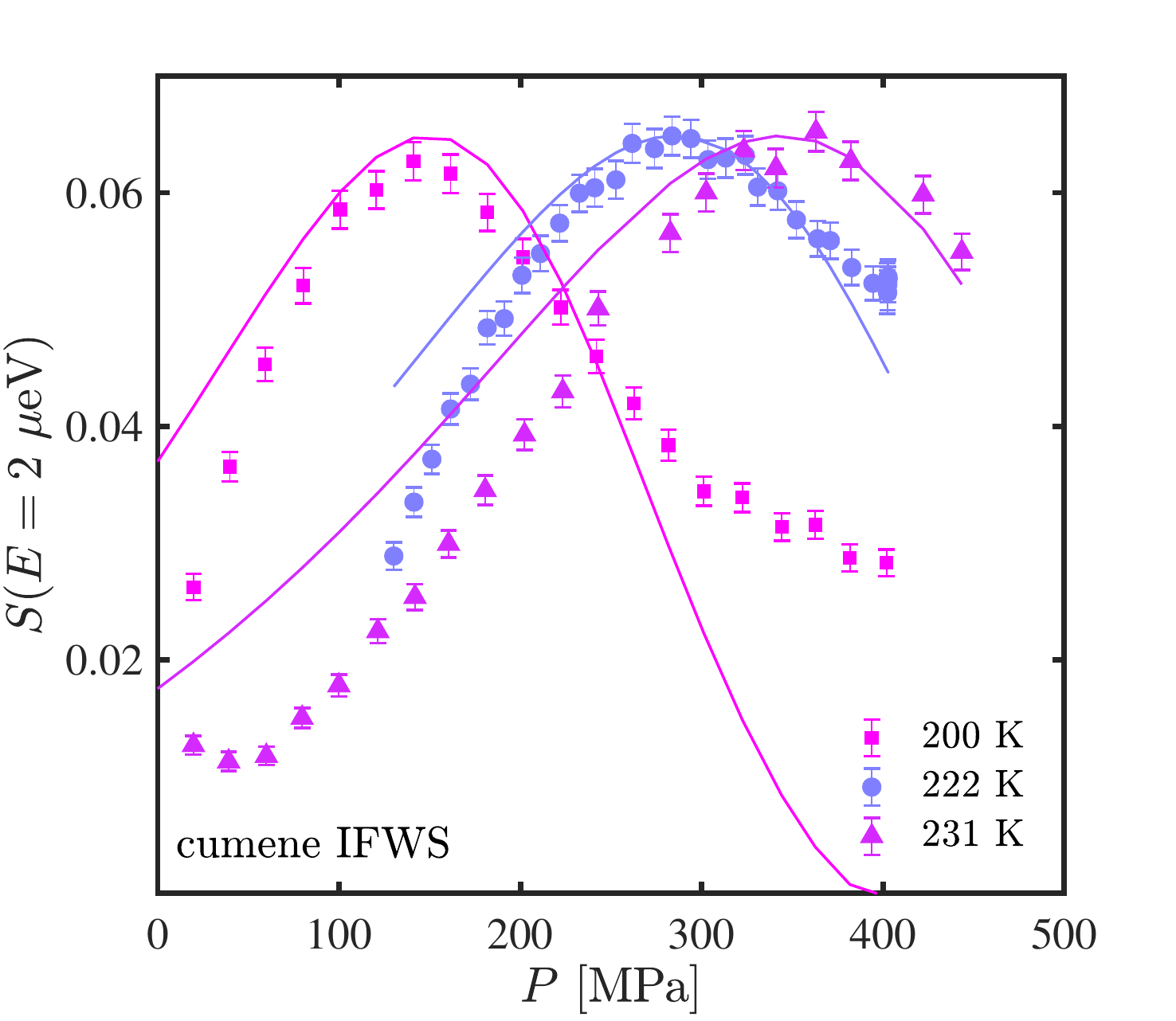}
\caption{Comparison of the simple model in Eq.~\ref{eq:res:ifws_model} (full lines) based on the relaxation times obtained from the full cumene spectra, and data on three isotherms from IFWS on IN16B. The density scaled IFWS data are shown to the left and as a function of pressure to the right.}\label{fig:res:cu_mct}
\end{figure}

Using the mode-coupling parameters reported in Ref.~\onlinecite{Li95}, and an expression for the temparature dependence of the relaxation time reported in Ref.~\onlinecite{Gotze87}, then substituting $\Gamma=\rho^\gamma/T$ for the temperature dependence and $\Gamma_c=\rho_c^\gamma/T_c$ for the critical temperature, we end up with an expression for the relaxation time:
\begin{equation}\label{eq:res:tau_mct}
\tau=\frac{\tau_0}{((\Gamma-\Gamma_c)/\Gamma_c)^n},
\end{equation}
where $n$ is the MCT critical exponent. Only the prefactor $\tau_0$ is fitted to data in the equilibrium liquid above the critical temperature, which is reported to be $\sim\SI{160}{\K}$ at ambient pressure. We use this expression of the $\alpha$-relaxation time and then suggest a simple model for our neutron data with a single relaxation process moving through the instrument window at a fixed energy, $\SI{2}{\micro\eV}$, similar to the IFWS from IN16B. If the relaxation process was an exponential process, this would correspond to a Lorentzian, we therefore add a phenomenological stretching to account for the stretched exponential behaviour.
The simple model is then given by
\begin{equation}\label{eq:res:ifws_model}
I\propto\frac{\tau^\beta}{1+(\omega^2\tau^2)^\beta}.
\end{equation}
In Fig.~\ref{fig:res:cu_mct}, we have used this model with the expression for the relaxation time as a function of $\rho^\gamma/T$ given in Eq.~\ref{eq:res:tau_mct} to recreate the IFWS shown as a function of $\rho^\gamma/T$ and finally, as a function of pressure. This model suggests only one distribution of relaxation times and is in qualitative agreement with our data, recreating the position of the maximum of the IFWS from the relaxation time obtained from fits to the full spectra and Eq.~\ref{eq:res:tau_mct}, thereby supporting our interpretation of the FWS.


\subsection{van der Waals liquid -- PPE}

\begin{figure}[htpb!]
\centering
\includegraphics[width=0.49\columnwidth]{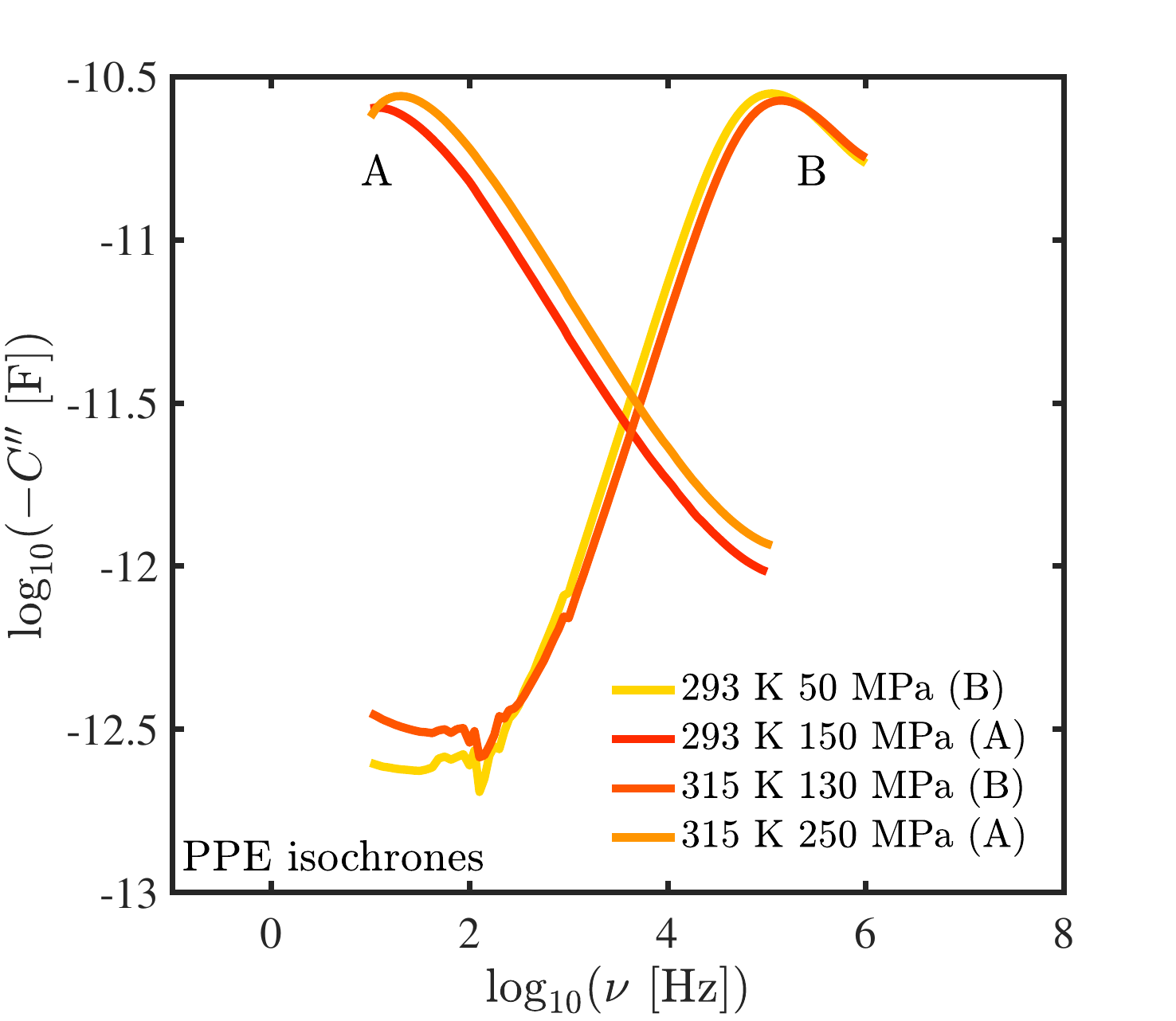}
\includegraphics[width=0.49\columnwidth]{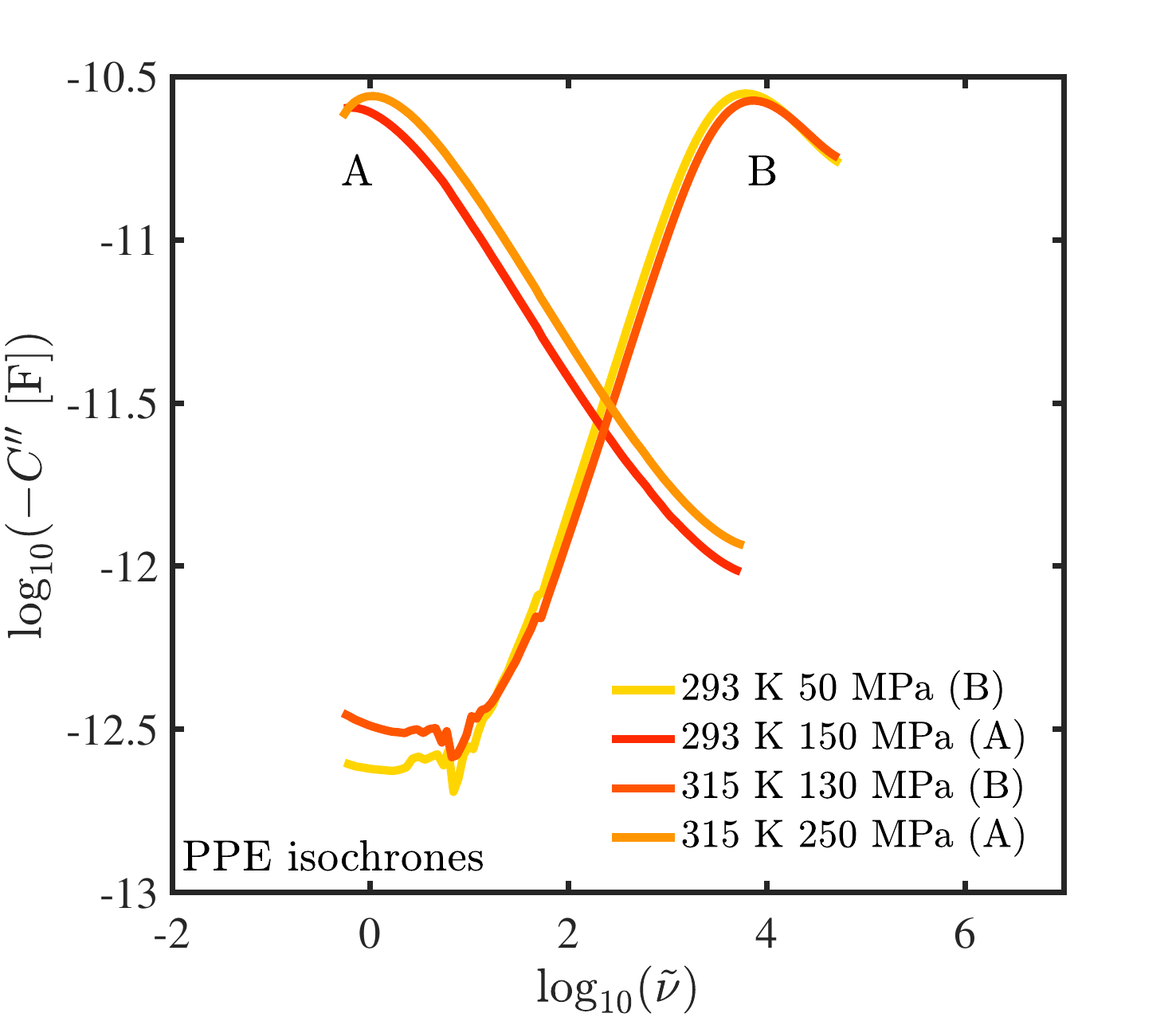}
\caption{Two pairs of isochronal state points determined from the dielectric loss peak on PPE measured on IN5, $\lambda=\SI{5}{\angstrom}$. Relaxation time for isochrone A is $\tau_\alpha\approx\SI{e-2}{\s}$ and for B $\tau_\alpha\approx\SI{e-6}{\s}$. Dielectric loss plotted as a function of frequency on an absolute scale (left) and as a function of the reduced frequency unit (right).}\label{fig:res:ppe_diel}
\end{figure}

The second van der Waals liquid is PPE, which only barely shows relaxation in the IN16B instrument window in the temperature and pressure range accessible with this setup. Instead, PPE data are reported from IN5 at \SI{5}{\angstrom} in the combined dielectric and neutron sample cell, i.e. accessing dynamics with neutrons on picosecond timescale. Two pairs of isochronal state points were found with high precision from the dielectric loss peak in the simultaneous dielectric and neutron measurements. The two pairs of isochronal state points are shown in Fig.~\ref{fig:res:ppe_diel} with relaxation times of roughly \SI{e-2}{\s} and \SI{e-6}{\s} for isochrone (A) and (B), respectively. The dielectric spectra are shown both as a function of frequency on an absolute frequency scale and in reduced frequency units. We observe that within the experimental uncertainty there is no influence on the relative position of the state points, i.e. isochronal state points on an absolute scale are also isochronal in reduced units. This observation supports the idea of identifying possible isomorphs by isochrones, an approach that was also taken studying the ultraviscous liquid in Ref.~\onlinecite{hwh18}.

\begin{figure}[htpb!]
\centering
\begin{minipage}{0.3\columnwidth}
\vspace{-7 em}
{\noindent PPE \\
\vspace{0.5 em} \noindent IN5, $\lambda=\SI{5}{\angstrom}$  \\
\vspace{0.5 em} \noindent $\sim\SI{e-11}{\s}$}
\end{minipage}
\includegraphics[width=0.32\columnwidth]{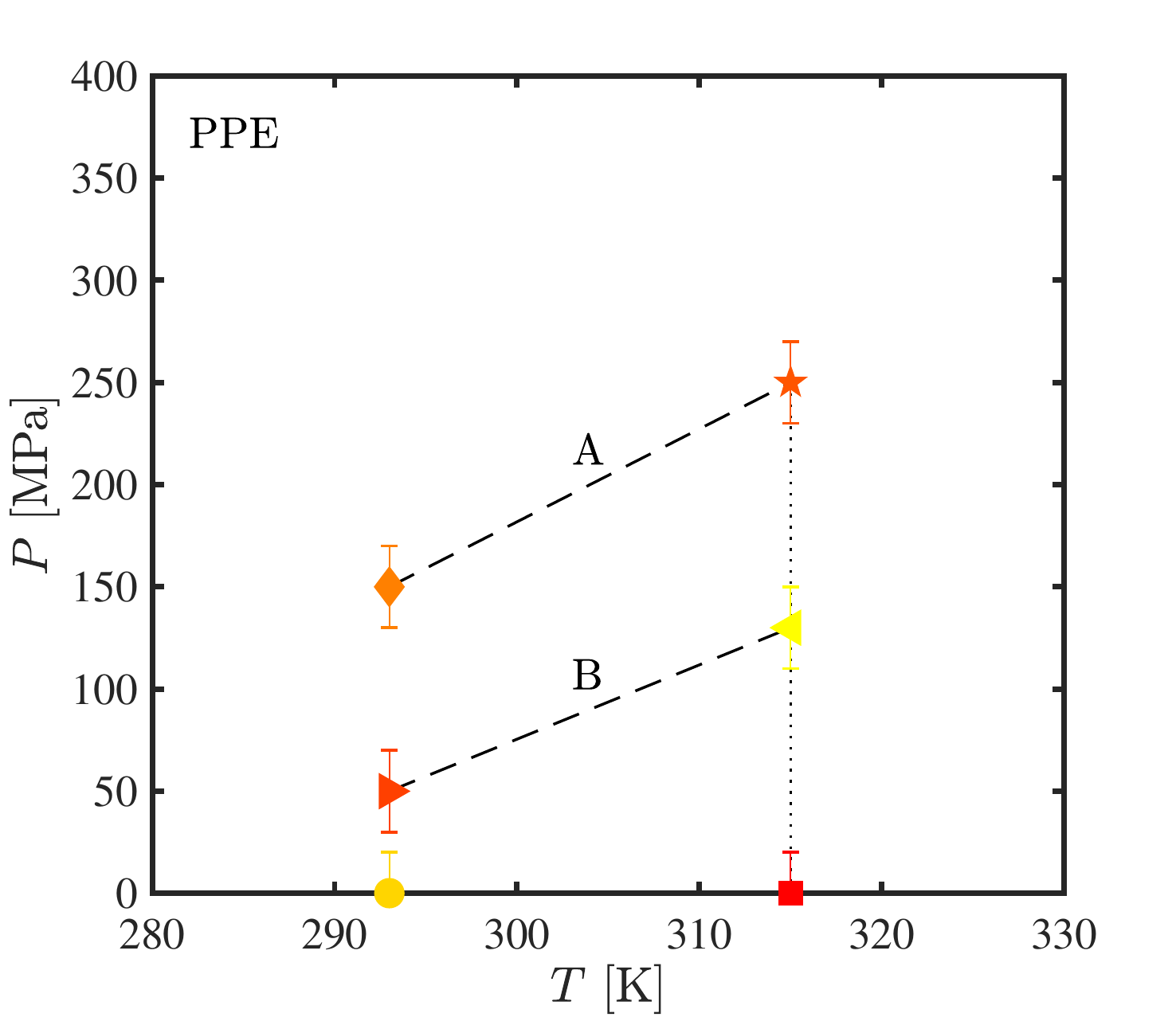} \\
\includegraphics[width=0.49\columnwidth]{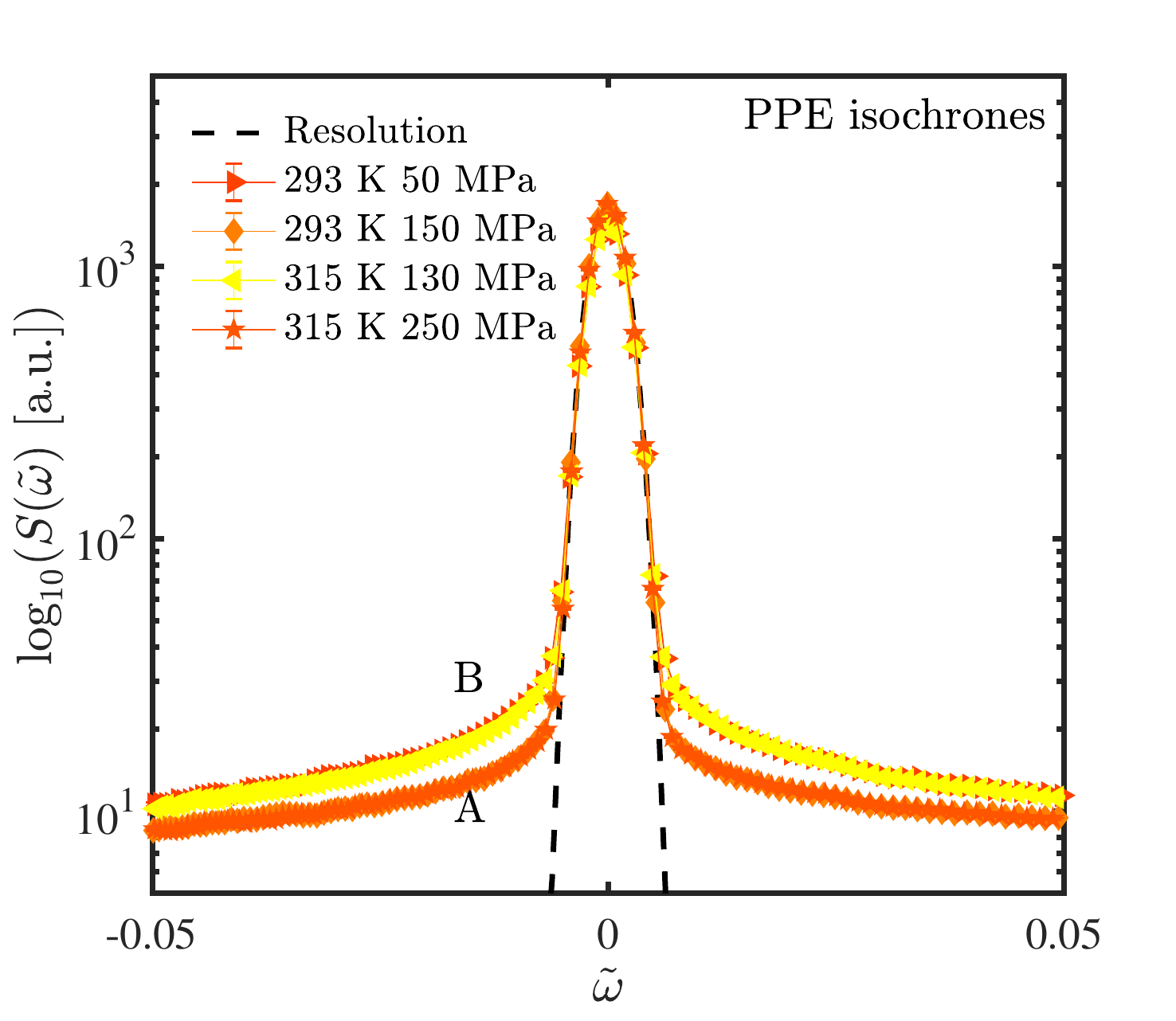}
\includegraphics[width=0.49\columnwidth]{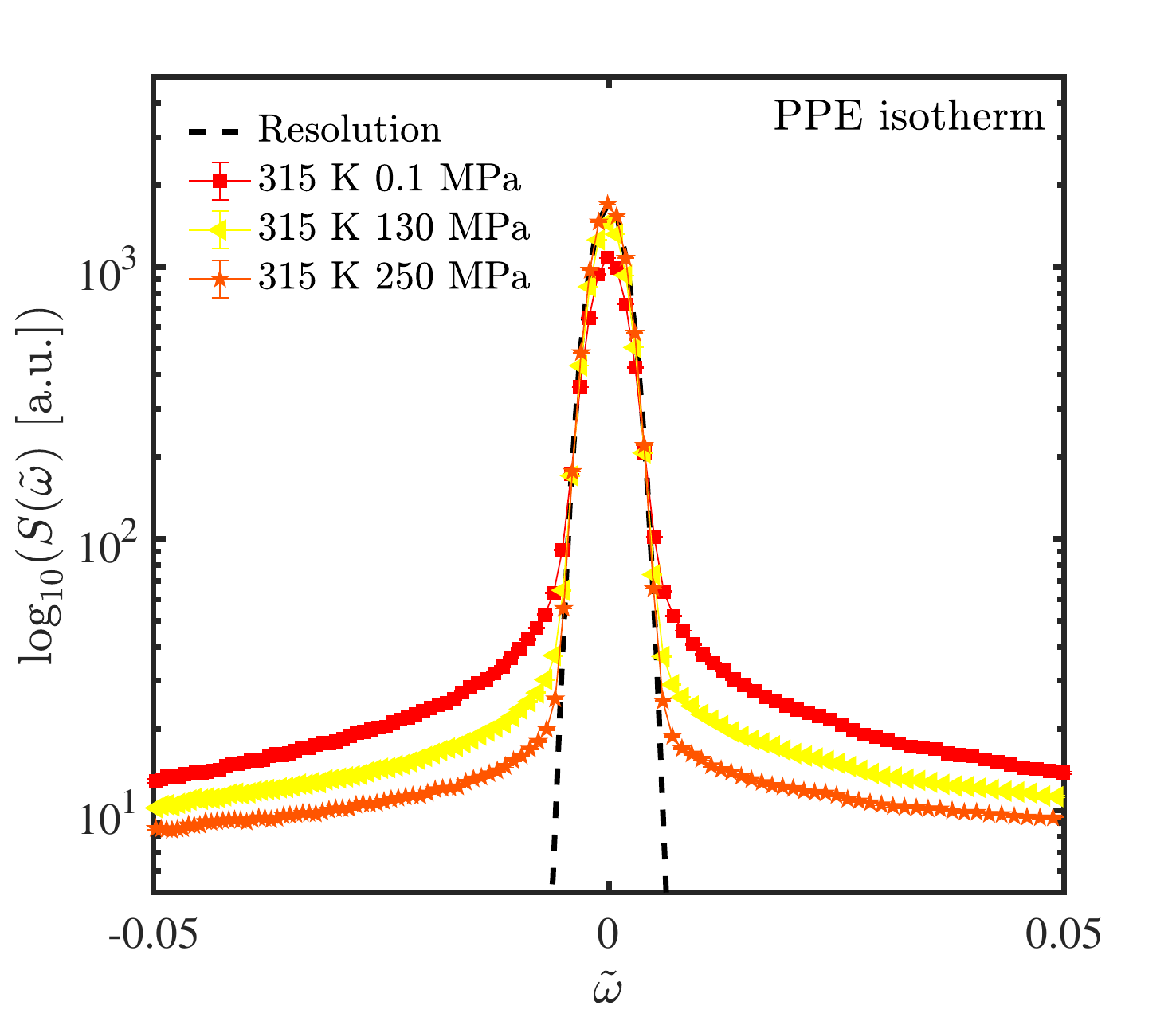}
\caption{Spectra on PPE from IN5 at $\lambda=\SI{5}{\angstrom}$ summed over $Q$. Top: The phase diagram shows the studied state points as a function of temperature and pressure and indicates the two pair of isochronal state points. Bottom: Two pairs of isochronal state points found from dielectrics (left) and an isotherm at \SI{315}{\K} for comparison (right).}\label{fig:res:ppe_spectra}
\end{figure}

The IN5 neutron spectra are shown in Fig.~\ref{fig:res:ppe_spectra}
for the two pairs of isochronal state points (A) and (B). The relaxation times of PPE
covered here on isochrones and isotherms are located between the
relaxation time found for cumene in the previous subsection and the
glass transition, a dynamic range not accessible in cumene because of crystallization. For the isotherm, we observe the dynamics slowing down as the liquid is compressed, i.e. less quasielastic broadening. For the isochronal state points, we observe an invariance of the spectral shape and the quasielastic broadening. This collapse is far from trivial; what we see on picosecond timescales is interpreted as the tail of the $\alpha$-relaxation stretching over many orders of magnitude, with vibrations and fast relaxation are here merged with the $\alpha$-relaxation\cite{hwh17}, suggesting that the whole relaxation curve is invariant for isochronal state points also in the liquid state \cite{hwh18}.

\subsection{Hydrogen-bonding liquid -- DPG}

\begin{figure}[htpb!]
\centering
\includegraphics[width=0.49\columnwidth]{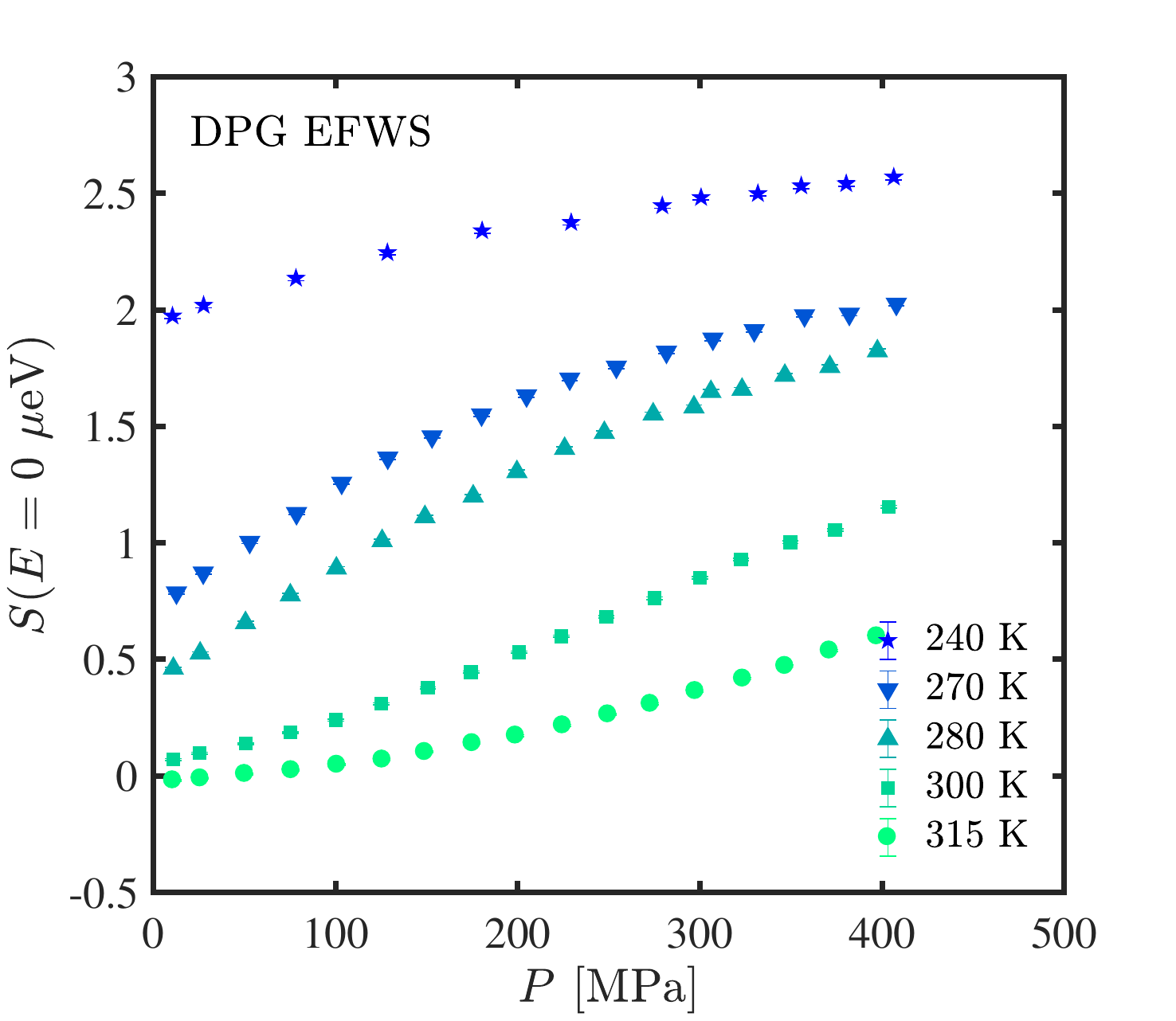}
\includegraphics[width=0.49\columnwidth]{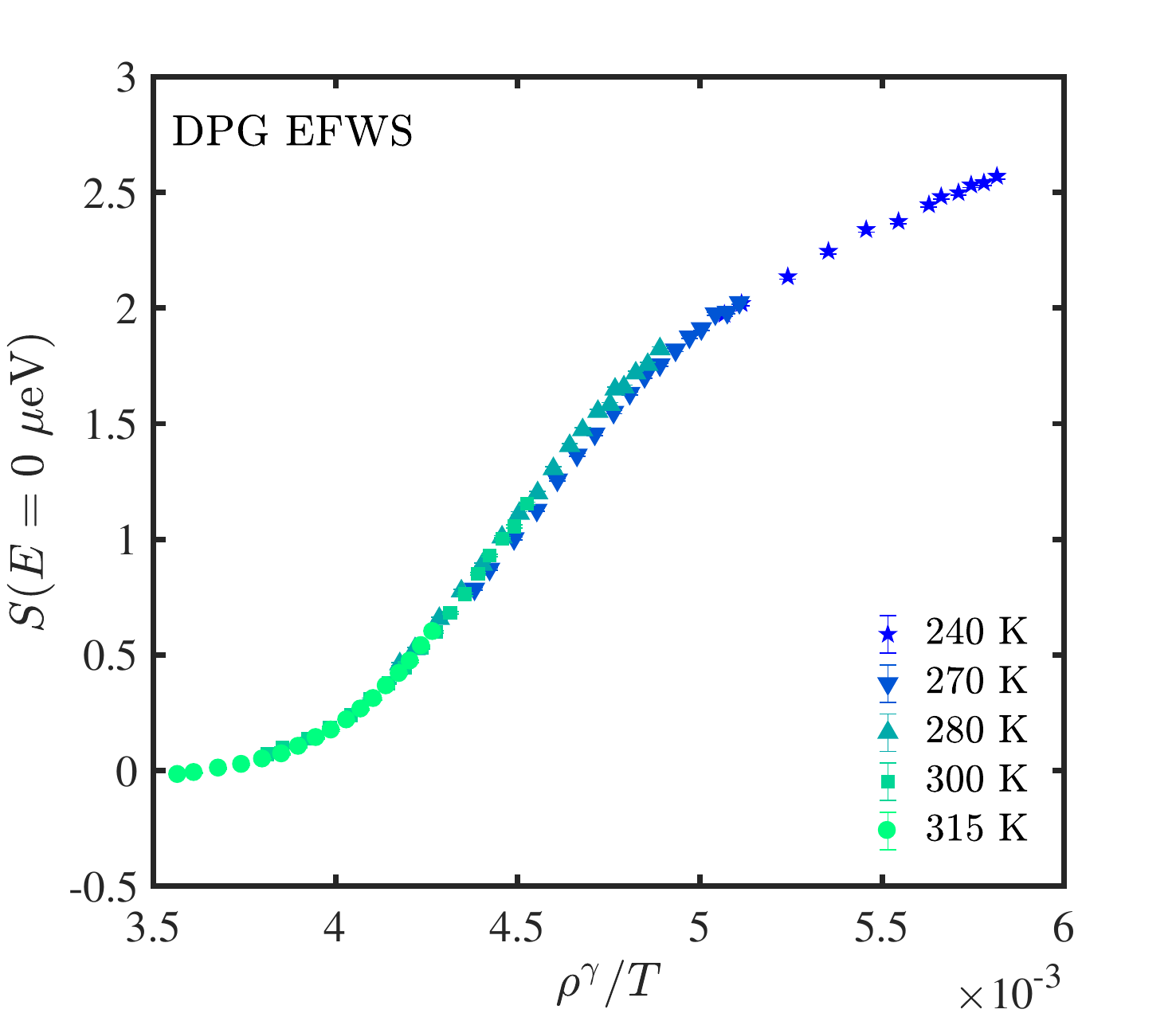} \\
\includegraphics[width=0.49\columnwidth]{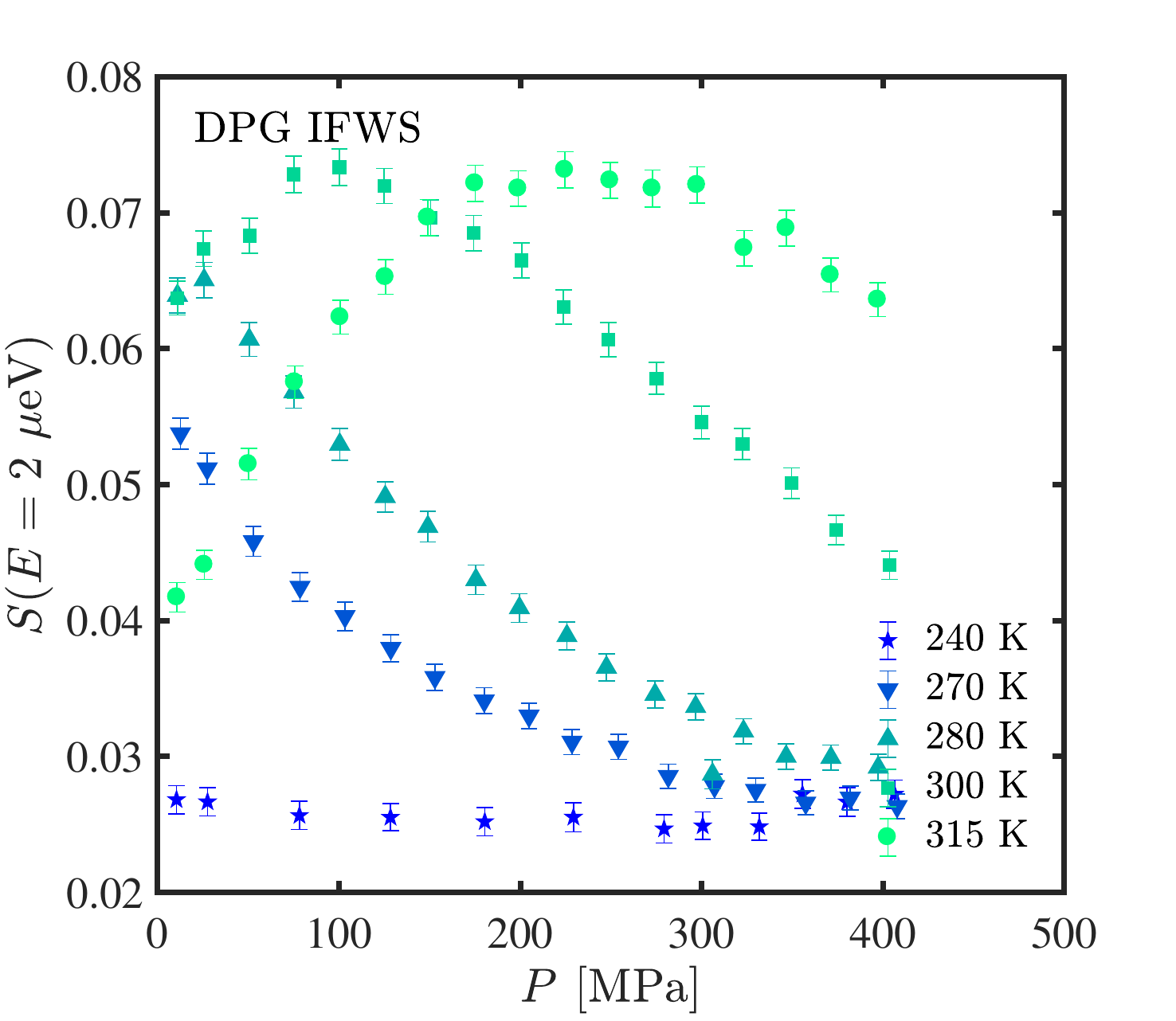}
\includegraphics[width=0.49\columnwidth]{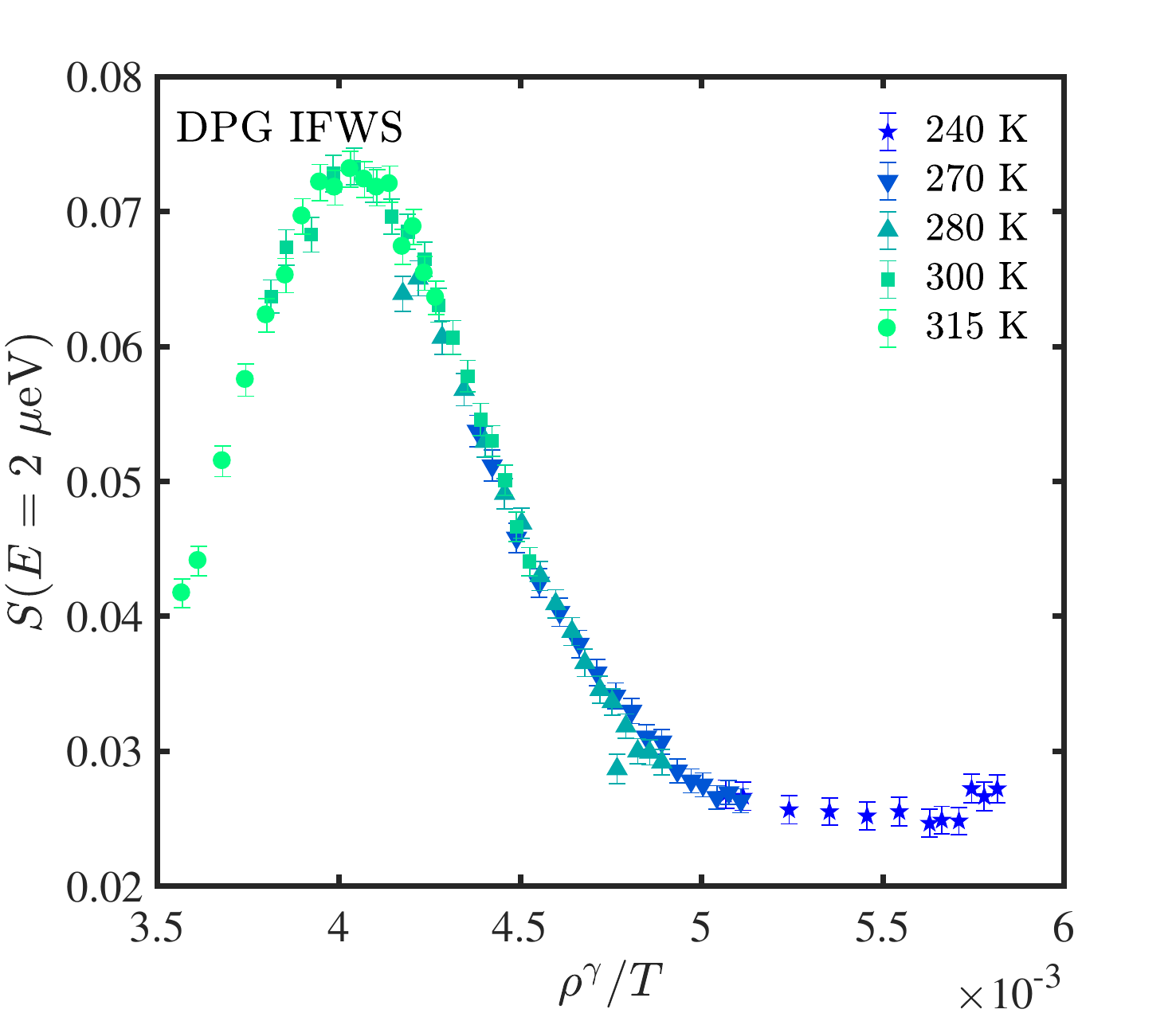} \\
\includegraphics[width=0.49\columnwidth]{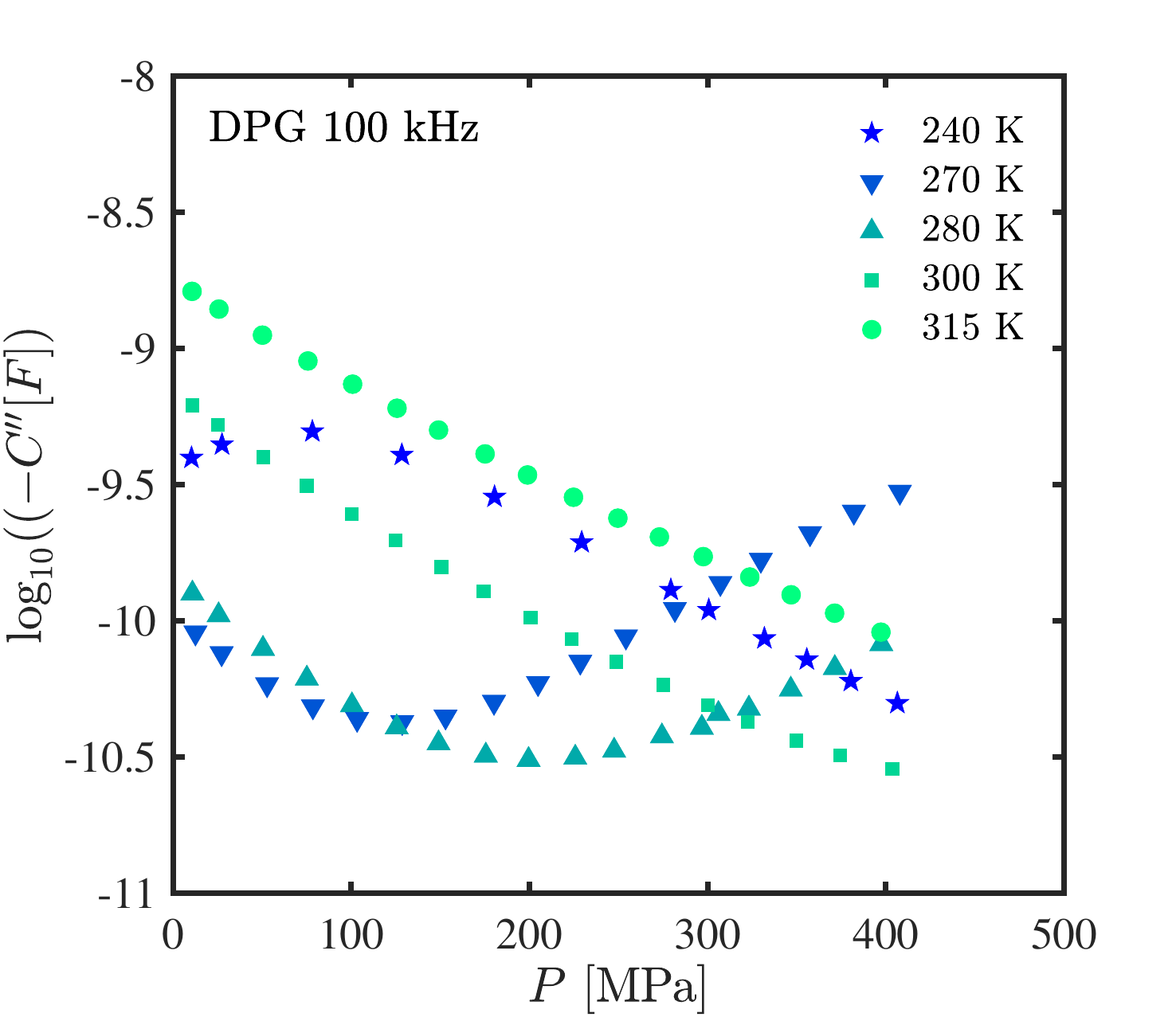}
\includegraphics[width=0.49\columnwidth]{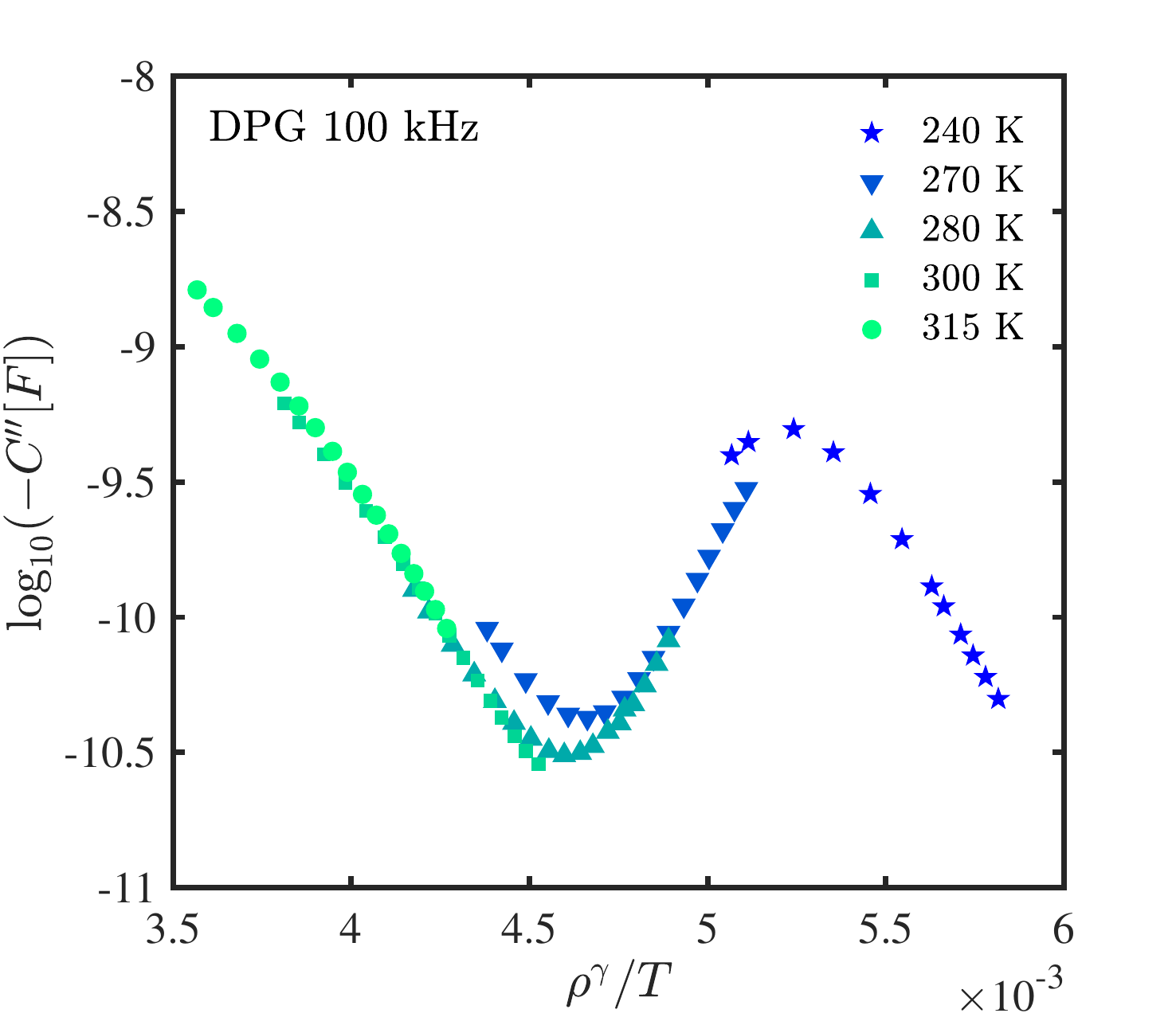}
\caption{FWS from IN16B on DPG summed over $Q$. Intensity of EFWS (top) and IFWS (middle) and fixed frequency dielectric loss (bottom) plotted for five isotherms (\SI{240}{\K}, \SI{270}{\K}, \SI{280}{\K}, \SI{300}{\K}, \SI{315}{\K}) as a function of pressure (left) and as a function of $\Gamma=\rho^\gamma/T$ (right).}\label{fig:res:dpg_fws}
\end{figure}

\begin{figure}[htpb!]
\centering
\begin{minipage}{0.3\columnwidth}
\vspace{-7 em}
{\noindent DPG \\
\vspace{0.5 em} \noindent IN16B \\
\vspace{0.5 em} \noindent $\sim\SI{e-9}{\s}$}
\end{minipage}
\includegraphics[width=0.32\columnwidth]{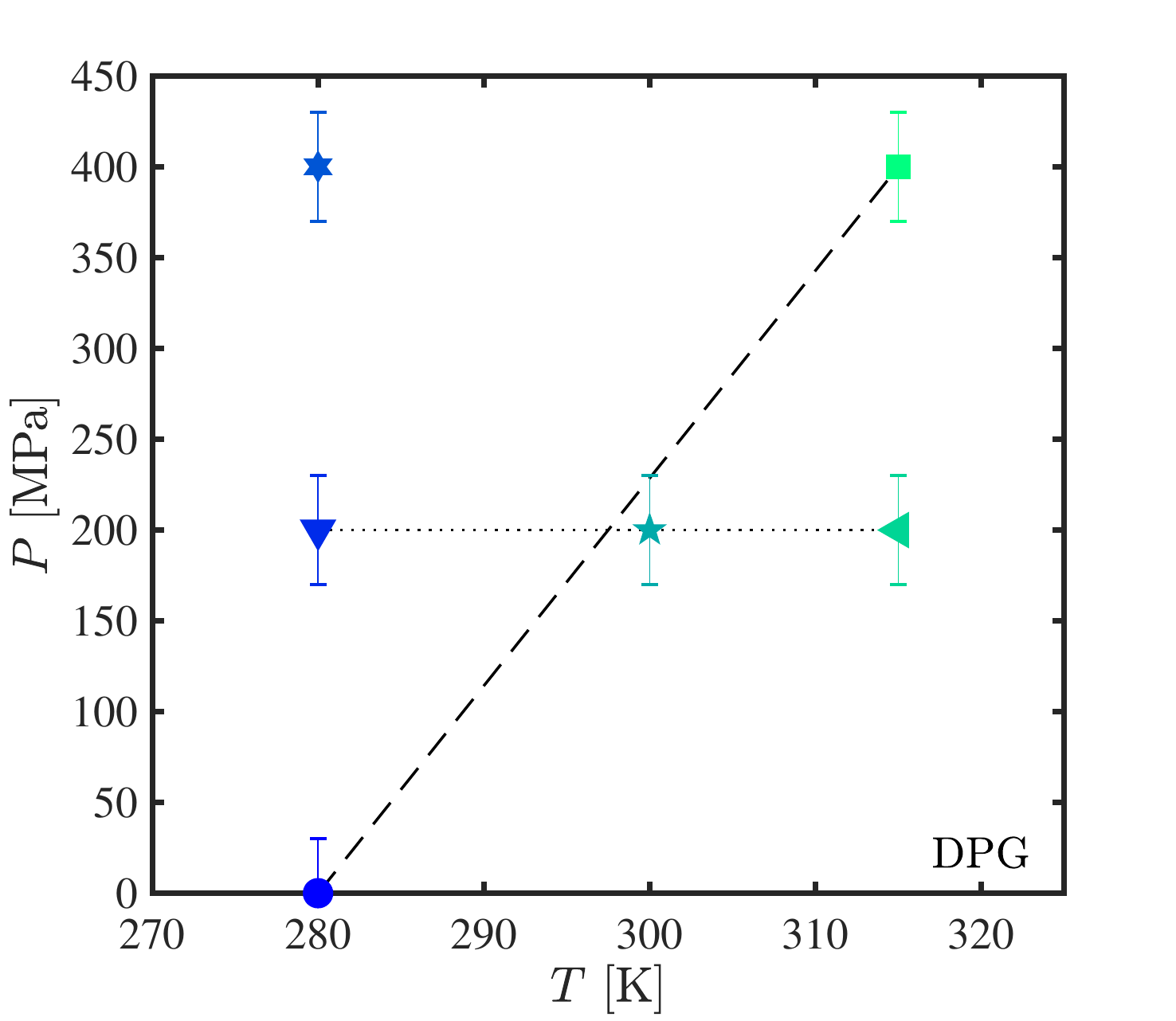} \\
\includegraphics[width=0.49\columnwidth]{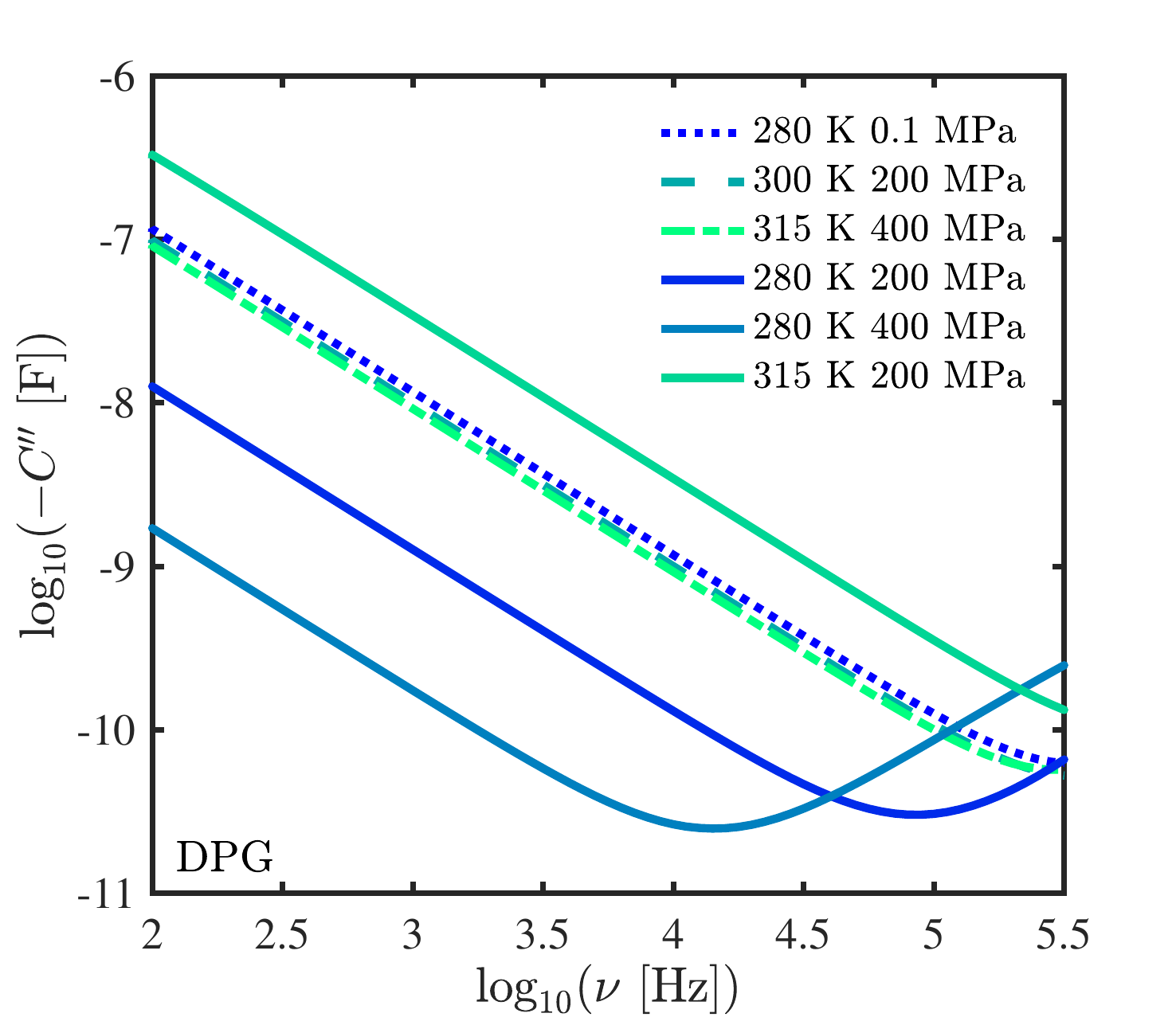} \\
\includegraphics[width=0.49\columnwidth]{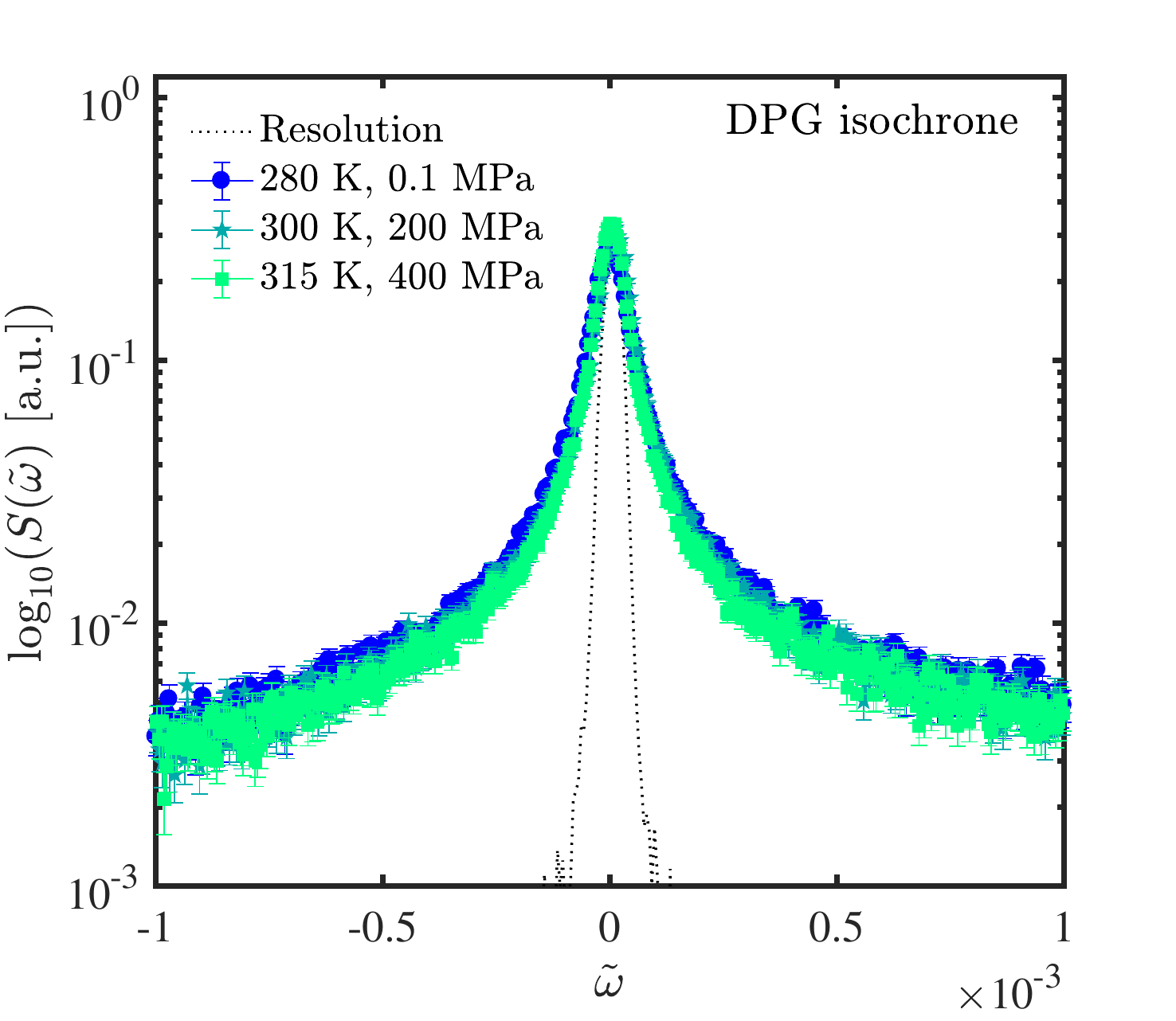}
\includegraphics[width=0.49\columnwidth]{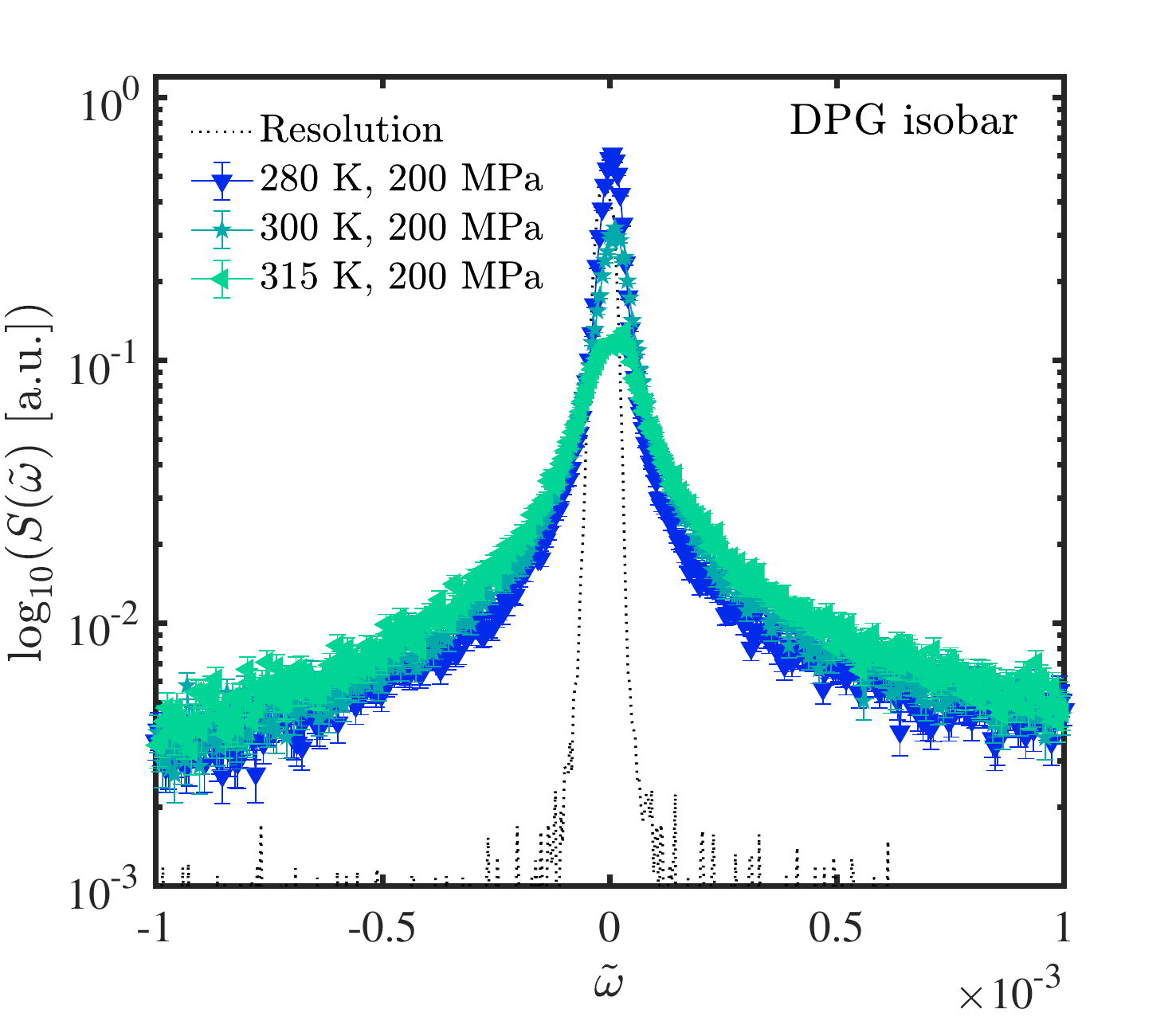}
\caption{Top: $(T,P)$-phase diagram with the studied state points of DPG on IN16B. Centre: dielectric loss as a function of frequency. Bottom: Spectra on DPG summed over $Q$ showing an isochrone (left) determined from FWS on IN16B and the dielectrics and an isobar for comparison (right).}\label{fig:res:dpg_spectra}
\end{figure}

The third set of data that we present is measured on the hydrogen-bonding liquid DPG at IN16B.
We present fixed window scans, full spectra and simultaneously acquired dielectric data, securing the precision of the measurements. In Fig.~\ref{fig:res:dpg_fws}, we show the measured EFWS and IFWS intensity at $\Delta E=\SI{0}{\micro\eV}$ and $\Delta E=\SI{2}{\micro\eV}$, respectively, as a function of pressure and as a function of $\rho^\gamma/T$, using $\gamma=1.5$, for five isotherms, and a similar plot with the dielectric loss signal at a fixed frequency of \SI{e5}{\Hz}. We observe an increase in the EFWS intensity upon compression as the mobility is decreased. From the IFWS, we observe that the high-temperature isotherms go through a maximum, similar to what was observed for cumene (Fig.~\ref{fig:res:cumene_fws}), i.e. the $\alpha$-relaxation passes through the neutron instrument window. Equivalently, for the low temperature isotherms, the $\alpha$-relaxation is observed to pass through the fixed frequency of the dielectrics. As the $\alpha$-relaxation moves out of the dielectric window, we observe DC-conductivity, also evident from the full frquency range in Fig.~\ref{fig:res:dpg_spectra}. We observe how the respective intensities can be made to collapse to a master curve for the dynamics on neutron nanosecond timescale when plotted as a function of $\rho^\gamma/T$
, and to some degree for the dielectric signal measured at roughly microsecond timescale.

 Just as for cumene, an isochrone was determined from the EFWS and IFWS intensity, but in this case also from the DC-conductivity and the wing of the relaxation observed in the dielectrics. A temperature-pressure phase diagram illustrating the studied state points is shown in Fig.~\ref{fig:res:dpg_spectra} along with the full spectra from IN16B and the dielectric signal at those state points. The isochrone determined from the fixed window neutron signal is observed also to be invariant in the dielectrics where mainly the DC-conductivity is observed and just the tail of the $\alpha$-relaxation at high frequencies. The full spectral shape and quasielastic broadening of the IN16B data is observed to be invariant for the isochrone. An isobar is shown for comparison, where we observe the dynamics speeding up as the temperature is increased, causing more broadening and less elastic intensity.

\begin{figure}[htpb!]
\centering
\includegraphics[width=0.49\columnwidth]{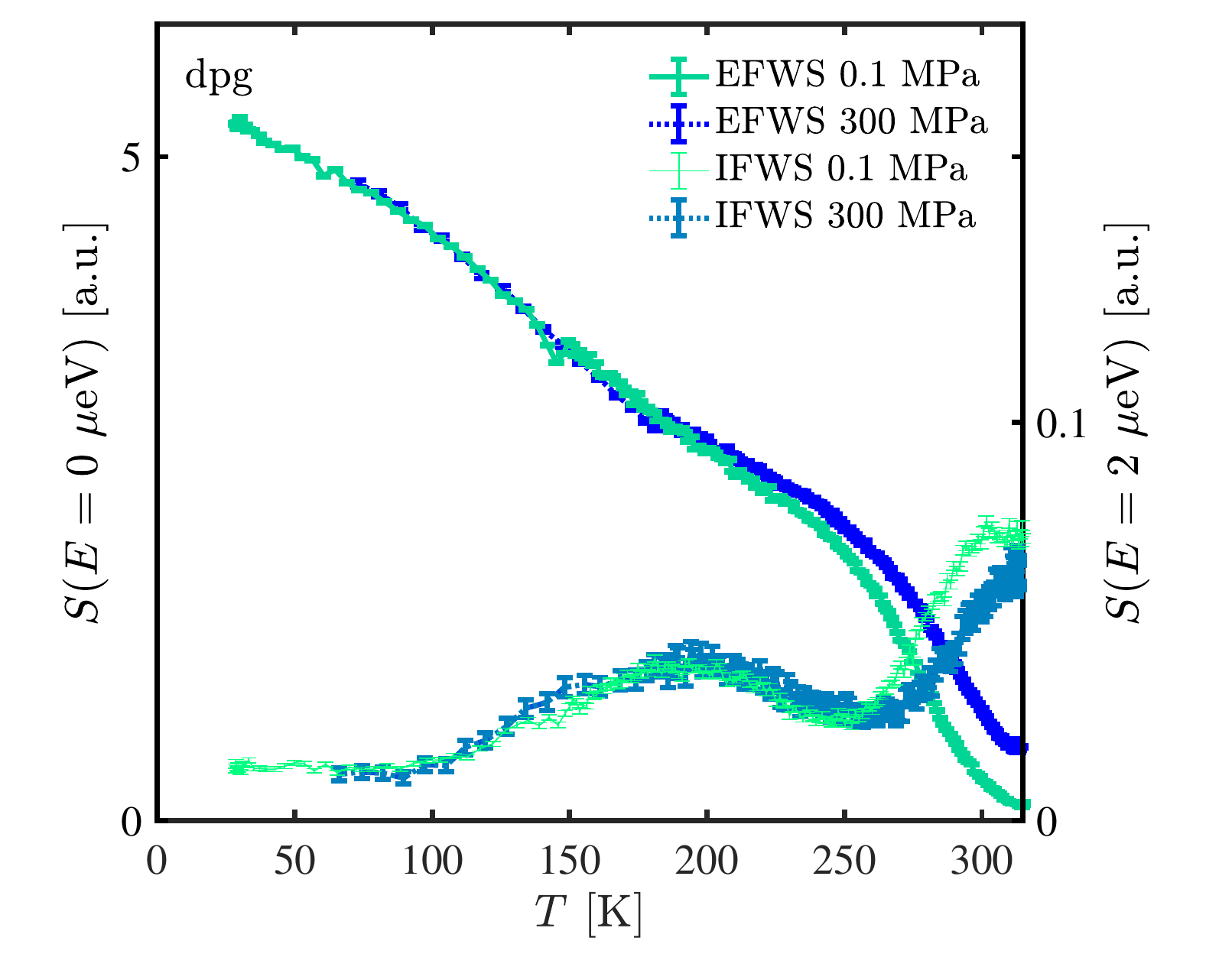} \\
\includegraphics[width=0.49\columnwidth]{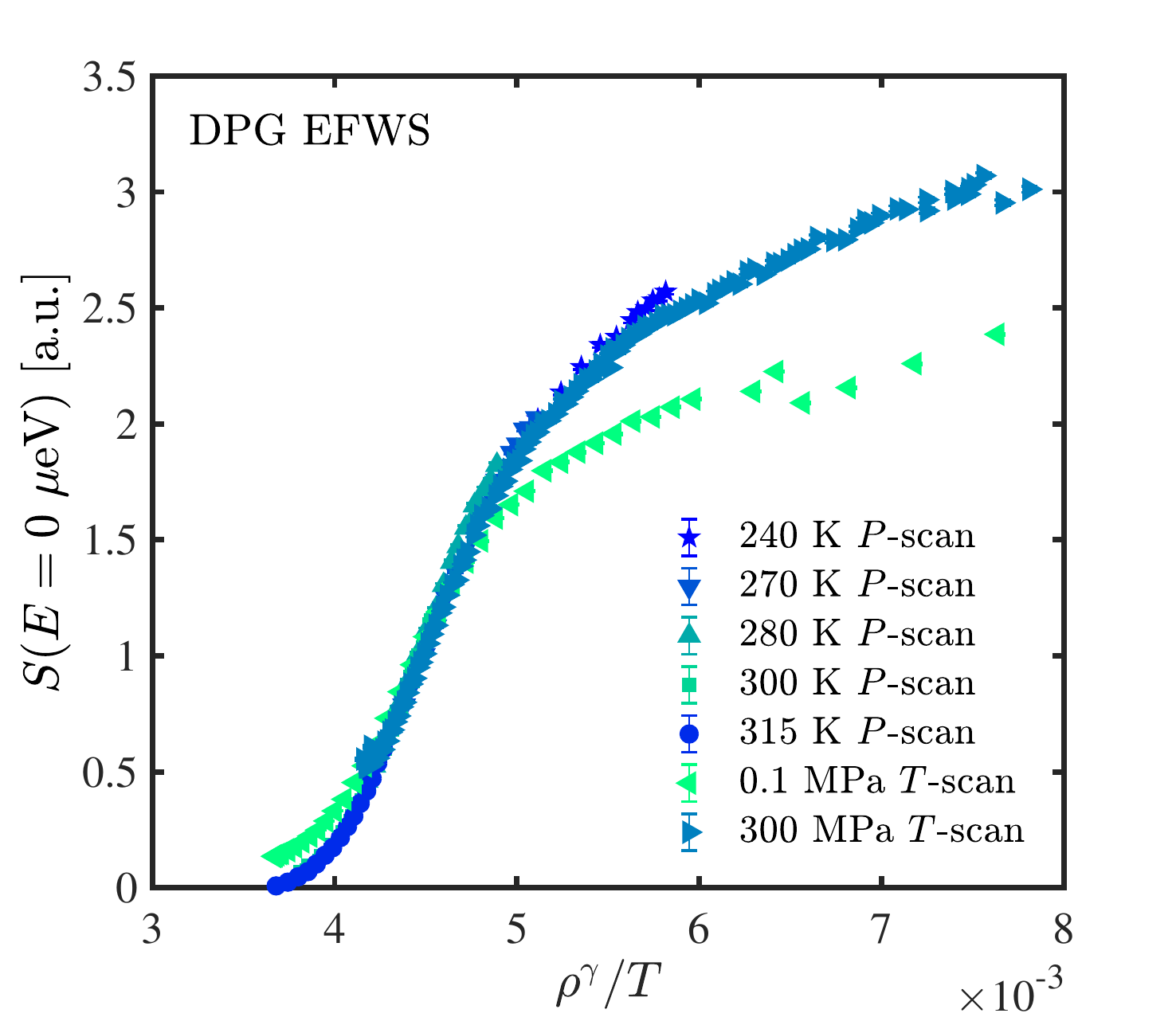}
\includegraphics[width=0.49\columnwidth]{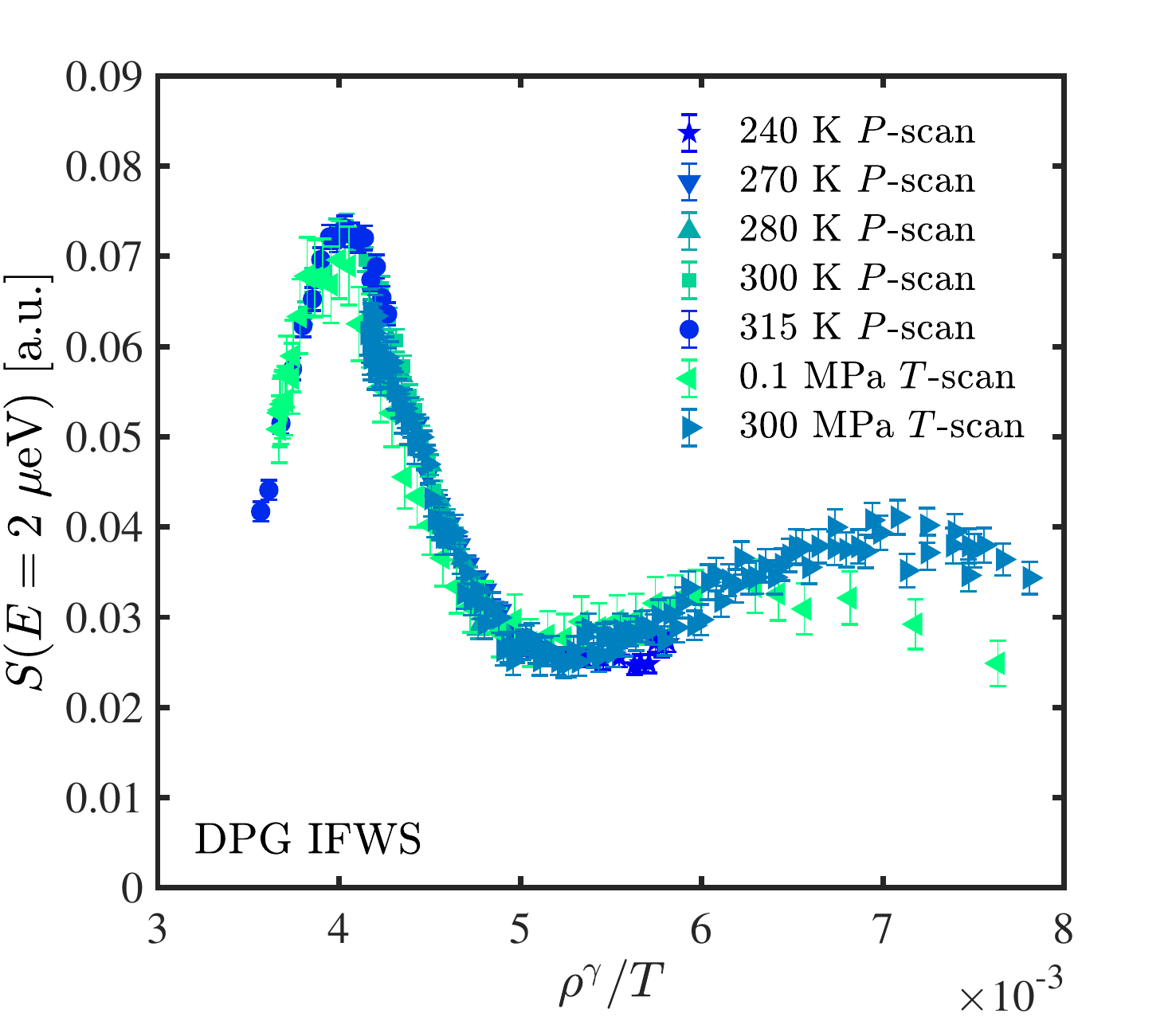}
\caption{Top: EFWS and IFWS from IN16B on DPG along two isobars at ambient pressure and \SI{300}{\mega\pascal}. Bottom: Density scaling of the EFWS (left) and IFWS (right) that are shown above and in Fig.~\ref{fig:res:dpg_fws}, i.e. for both the isobars and isotherms.}\label{fig:res:dpg_TP}
\end{figure}

\begin{figure}[htpb!]
\centering
\includegraphics[width=0.49\columnwidth]{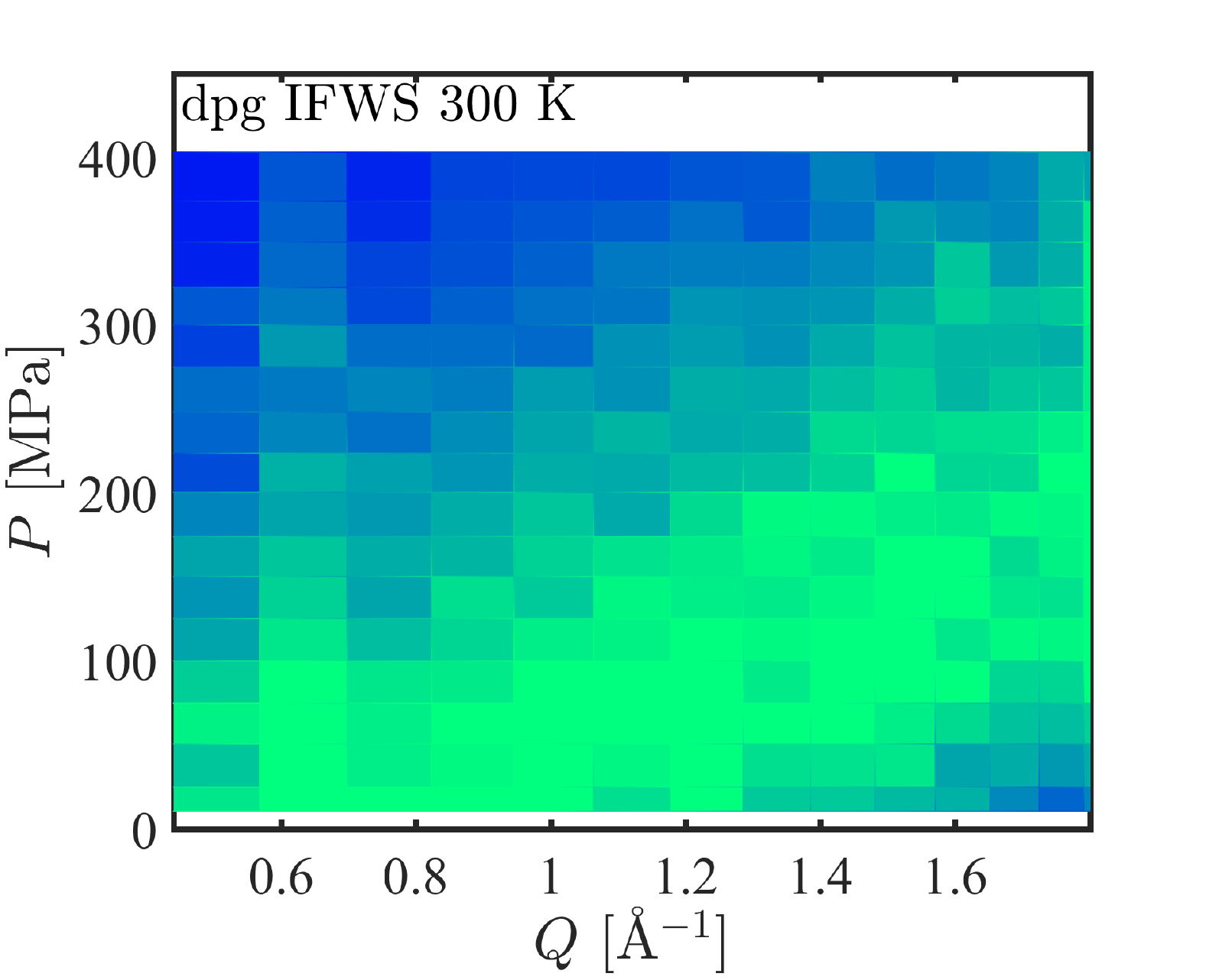}
\includegraphics[width=0.49\columnwidth]{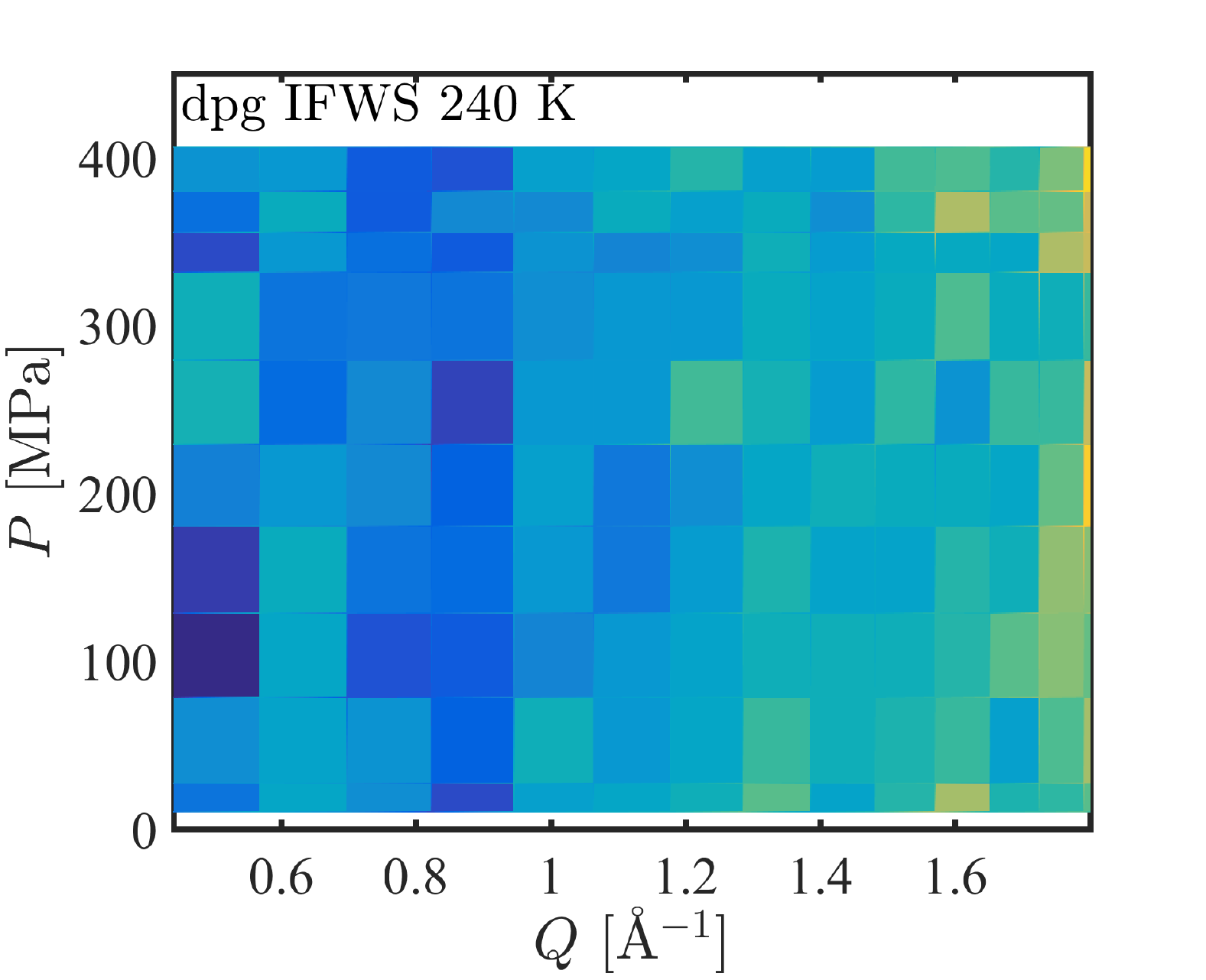}
\caption{$Q$-dependence of the IFWS of two isotherms from IN16B on DPG at \SI{300}{\K} (left) and \SI{240}{\K} (right).}\label{fig:res:dpg_Q}
\end{figure}

The spectra from IN16B shown in Fig.~\ref{fig:res:dpg_spectra} are all from state points where the $\alpha$-relaxation is near the timescale of the backscattering instrument window of IN16B, i.e. nanoseconds. In Fig.~\ref{fig:res:dpg_TP} (top), we present fixed window scans from two isobars, at ambient pressure and at \SI{300}{\mega\pascal}, carried out from the high temperature region, where the $\alpha$-relaxation is in the instrument window, and all the way into the glassy state. On cooling at IN16B, DPG is observed to have a distinct signal with a maximum close to the glass transition, $T_g(P_\mathrm{atm})=\SI{195}{\K}$. This signal is well separated from the $\alpha$-relaxation on IN16B which is roughly \SI{100}{\K} higher at ambient pressure. While the $\alpha$-relaxation with a maximum at $\sim\SI{300}{\K}$ is observed to move upon increased pressure, the second bumb at lower temperatures does not respond to pressure, also visible in the pressure scans in Fig.~\ref{fig:res:dpg_fws}, which indicates that this contribution is from intra-molecular dynamics.

If we compare the $Q$-dependence for the isotherm at \SI{240}{\K} to that at \SI{300}{\K}
(Fig.~\ref{fig:res:dpg_Q}), we see that for the high temperature isotherm, there is a clear shift on compression in the relaxation process towards higher $Q$, indicating that this is of translational character, whereas there is no $Q$-dependence observed for the isotherm at \SI{240}{\K}, suggesting a local process with a higher intensity at higher $Q$. This signal most likely stems from the methyl-group rotation of DPG and is manifested in the IFWS as the second broad bump at lower temperatures than the $\alpha$-relaxation with a maximum around \SI{200}{\K} (Fig.~\ref{fig:res:dpg_TP}). By combining the two isobars with the five isotherms in Fig.~\ref{fig:res:dpg_fws}, we observe that when the EFWS and IFWS are plotted as a function of $\rho^\gamma/T$, density scaling breaks down for the methyl-group rotation dynamics and the dynamics can no longer be expressed solely by $\Gamma=\rho^\gamma/T$.



\section{Discussion}

In the previous section, we showed how dynamics of the $\alpha$-relaxation taking place on timescales accessible with neutron and dielectric spectroscopy can be made to collapse into one curve when performing density scaling in the equilibrium liquid. The dynamics involved covers timescales from picoseconds to milliseconds and suggests that the timescale of the $\alpha$-relaxation is determined by the control parameter $\Gamma=\rho^\gamma/T$. We have shown isochronal superposition over several decades in time accessed by different spectrometers for the simple van der Waals liquid PPE, i.e. an invariance in the quasielastic broadening on picosecond timescale along isochrones determined from dielectrics. Also for DPG, we observed that the DC-conductivity visible from dielectrics and the full spectral shape and linewidth were invariant along the isochrone determined from the FWS on IN16B and the dielectric spectra. For cumene, that was measured on two different spectrometers and combined to cover almost four orders of magnitude in the time domain, allowed for a more careful analysis of the relaxaton.

The spectral shape of the Fourier-transformed cumene data was fitted with the same stretching exponent $\beta=0.5$ for all temperatures and pressure, which suggests time-temperature-pressure superposition, i.e. that the shape of the $\alpha$-relaxation is always the same in the equilibrium liquid far from the glass transition. This observation implies an extra level of simplicity that seems to apply to some liquids that makes the observation of isochronal superposition an evident observation.

Most density scaling studies have been performed in the viscous liquid close to the glass transition. In this study, we move away from the slow-flowing liquid and into the liquid state. The $\alpha$-relaxation times for cumene and DPG were in this study mainly in the instrument window of IN16B, i.e. on nanosecond timescale. Density scaling for the $\alpha$-relaxation dynamics was shown for the two liquids with very different values of $\gamma$, both reported elsewhere and found in different dynamic range measured with different techniques. While the scaling exponent $\gamma$ is allowed to vary, we observe that for cumene $\gamma=4.8$ (Ref.~\onlinecite{Ransom17}) works well within the experimental uncertainty that we have to work with. The $\gamma$-value of cumene is fairly high, meaning that density plays a much larger role for the dynamics, as compared to the hydrogen-bonding liquid DPG, where the scaling exponent is just $\gamma=1.5$ (Ref.~\onlinecite{Grzybowski06_gamma}), meaning that the dynamics therefore mainly is affected by temperature.

In agreement with previous observations \cite{Adrjanowicz16,Puosi16,Romanini17}, we observe density scaling to work well also for the tested hydrogen-bonding liquid when it comes to the $\alpha$-relaxation. But we observed density scaling to break down for the intra-molecular dynamics of the methyl-group rotation that was identified from the lack of pressure response and the $Q$-dependent signal. Previously, we have observed isochronal superposition to break down for the hydrogen-bonding liquid DPG along the glass transition when fast relaxation and vibrations were completely separated from the $\alpha$-relaxation \cite{hwh18}.

The fact that we observe density scaling to break down in this study for the intra-molecular motion in DPG is in agreement with observations from molecular dynamics simulations \cite{Veldhorst15,Olsen16}. Spring-like bonds were introduced as intra-molecular motion and what was referred to as pseudo-isomorphs were recovered when only the intermolecular dynamics were treated, i.e. the scaling only worked for intermolecular motion. A partly deuterated DPG sample that would mask the methyl-group rotation in the neutron signal would therefore be an obvious next approach to test whether density scaling in this way can be restored for the full dynamic range.

\subsection{Acknowledgements}
This work was supported by the Danish Council for Independent Research (Sapere Aude: Starting Grant).
We acknowledge the workshop at IMFUFA and everyone involved in the LTP-6-7 at the ILL for technical support: instrument responsibles Judith Peters on IN13 and Jacques Olivier on IN5; Eddy Leli\'{e}vre-Berna, cryogenics and high-pressure from the SANE group.

\bibliographystyle{apsrev4-1}

\end{document}